\documentclass[12pt,usenames,dvipsnames,svgnames]{article}

\newcommand{\bl}{{\text{\sc b-l}}} %
\newcommand{\dtheta}{\dot{\theta}} %

\usepackage[square,comma,sort&compress,numbers]{natbib} %

\usepackage{booktabs}

\usepackage{graphicx,times,units} %
\usepackage{fullpage,subfig,slashed}%
\usepackage[thicklines]{cancel} 
\usepackage{amsmath,amssymb,amscd} %
\usepackage{esint} %

\usepackage[a4paper,hscale=.75,vscale=.8,centering]{geometry}
\usepackage{fullpage}

\usepackage{bm,paralist,xspace,url,relsize} %

\newcommand{\0}{{(0)}}%
\newcommand{\1}{{(1)}}%

\usepackage[colorlinks=true,bookmarks=false]{hyperref}

\newcommand{\tr}{{\mathrm{tr}}} 
\newcommand{\const}{\mathrm{const}} %
\newcommand{\parfrac}[2]{\left(\frac{#1}{#2}\right)}
\newcommand{\pfrac}[2]{\frac{\partial #1}{\partial #2}}
\newcommand{\CH}{\mathcal{H}} %
\newcommand{\CO}{\mathcal{O}} %
\newcommand{\axion}{\text{axion}} 

\begin{document}

\title{Magnetohydrodynamics of Chiral Relativistic Fluids}

\author{Alexey~Boyarsky$^1$, J\"{u}rg Fr\"{o}hlich$^2$, Oleg Ruchayskiy$^3$\\
  \small $^1$Instituut-Lorentz for Theoretical Physics, Universiteit Leiden\\
  \small Niels Bohrweg 2, Leiden, The Netherlands\\
  \small $^2$Institute of Theoretical Physics, ETH Zurich\\
  \small CH-8093
  Zurich, Switzerland\\
  \small $^3$Ecole Polytechnique F\'ed\'erale de Lausanne\\
  \small FSB/ITP/LPPC, BSP
  720, CH-1015, Lausanne, Switzerland}%

\date{}
\maketitle

\begin{abstract}
  \textit{We study the dynamics of a plasma of charged relativistic fermions
    at very high temperature $T\gg m$, where $m$ is the fermion mass, coupled
    to the electromagnetic field.  In particular, we derive a
    magneto-hydrodynamical description of the evolution of such a plasma. We
    show that, as compared to conventional MHD for a plasma of
    non-relativistic particles, the hydrodynamical description of the
    relativistic plasma involves new degrees of freedom described by a
    pseudo-scalar field originating in a local asymmetry in the densities of
    left-handed and right-handed fermions. This field can be interpreted as 
    an effective axion
    field. 
  Taking into account the chiral anomaly we present dynamical equations for 
  the evolution of this field, as well as of 
  other fields appearing in the MHD description of the plasma.  Due to its
  non-linear coupling to helical magnetic fields, the axion field
  significantly affects the dynamics of a magnetized plasma and can give rise 
  to a novel type of inverse cascade.}
\end{abstract}
 
\renewcommand{\abstractname}{}

\section{Introduction}
\label{sec:introduction}

It has been suggested ~\cite{Frohlich:2000en,Frohlich:2002fg}, more than a
decade ago, that the chiral anomaly may be accompanied by the appearance of some
\emph{new degrees of freedom} described by an effective axion field that
couples to the index density of the electromagnetic field.  More recently, it
has been understood{~\cite{Boyarsky:11a}} that this axion field plays an
important role in the dynamics of the electromagnetic field in a relativistic
plasma of charged fermions, (e.g., such as a plasma described by
 the Standard Model at high temperature
and/or large matter density).  In the spatially homogeneous case, the time
derivative of the axion field is proportional to the difference between the
chemical potentials (densities) of left-handed and right-handed charged
particles~\cite{Frohlich:2000en,Frohlich:2002fg}.  As a consequence of the
chiral anomaly, a left-right asymmetry, as described by a non-vanishing
time-dependent axion field, affects the dynamics of the electromagnetic field
(see also ~\cite{Vilenkin:80a,Joyce:97,Redlich:1984md,Alekseev:98a}) and gives
rise to an instability of solutions of the coupled Maxwell-axion equations
~\cite{Frohlich:2000en}.

This kind of instability is encountered in different physical systems
exhibiting a left-right asymmetry. For example, it has recently also appeared
in the electrodynamics of topological insulators; (see, e.g.,
\cite{Ooguri:11}). Instabilities leading to the growth of helical magnetic
fields in the plasma of relativistic charged particles, as discussed in this
paper, are expected to have played an important role in the dynamics of the
early Universe, assuming that a potent source of chiral asymmetry was
present;~(see, e.g.,
\cite{Joyce:97,Frohlich:2000en,Boyarsky:11a,Semikoz:11a,Semikoz:12a,Tashiro:12}).
Magnetic fields pervading the primordial plasma at early times would have had
significant effects on many processes, such as
baryogenesis~\cite{Giovannini:1997eg}, primordial
nucleosynthesis~\cite{Cheng:1996yi,Yamazaki:2012jd}, CMB physics~(see
\cite{Durrer:13} for a review), the growth of various inhomogeneities
(see~\cite{Barrow:2006ch}, and refs.\ therein), etc.  Thus, understanding the
dynamics of primordial magnetic fields in the presence of sources of
left-right asymmetries, as described by time-dependent effective axion fields,
is essential in attempts to describe the evolution of the early Universe.

 An important question in cosmology is whether traces of primordial magnetic
 fields could have served as seeds for galactic magnetic
 fields~\cite{Widrow:2002ud,Beck:2000dc} or given rise to the large-scale
 magnetic fields observed in
 voids~\cite{Neronov:10a,Tavecchio:10,Dolag:10}. The astrophysical origin of
 the magnetic fields in voids is still under active investigation; (see
 e.g.~\cite{Beck:13a,Beck:13b}). If their origin turned out to be primordial
 then the difficulty that causally independent homogeneous magnetic fields
 survive ~\cite{Banerjee:04,Durrer:2003ja,Caprini:2009pr} would have to be
 faced. To cope with this difficulty one might argue that the mechanism of
 generation of those fields is ``super-horizon'', presumably related to
 inflation ~\cite{Ratra:1991bn,Giovannini:2000dj}. However, the generation and
 survival of somewhat sizeable magnetic fields during inflation is difficult
 to reconcile with slow-roll
 conditions~\cite{Giovannini:2000wta,Demozzi:2009fu,Durrer:2010mq,Martin:2007ue}.

 In an expanding Universe filled with Standard-Model particles, chirality
 flipping reactions, which erase chiral asymmetries, are in thermal
 equilibrium at temperatures below $\sim80$ TeV~\cite{Campbell:92}. It has
 therefore been expected that an initial growth of magnetic fields, as derived
 from the chiral anomaly, could only have taken place at such truly enormous
 temperatures (if they ever existed in the early Universe). In an expanding
 Universe, the wavelength of any perturbation grows like the scale factor
 $a(t) \propto t^\alpha$ (with $\alpha < 1$, for any non-inflationary
 expansion), and the distance, $R_H$, to the horizon in the Universe increases
 linearly in time, $R_H \propto t$. As a result, even if a magnetic field were
 generated during some very early epoch, with correlations extending over a
 horizon-size scale (maximal scale possible for any causal generation
 mechanism), it would soon become stronger at sub-horizon scales. But it is
 known that sub-horizon-scale magnetic fields die out because of dissipation
 caused by resistivity and viscosity
 effects~\cite{Jedamzik:96,Subramanian:97,Banerjee:03,Banerjee:04,Giovannini:2003yn,Kahniashvili:2010gp}.
 It has been argued that the only way for causally generated magnetic fields
 to survive is the turbulence-driven inverse
 cascade~\cite{Christensson:2002xu,Banerjee:04,Durrer:2003ja,Campanelli:07} of
 helical magnetic fields -- a process that makes the characteristic scale of
 the magnetic fields grow \emph{faster} than red-shifting does. At present,
 the efficiency of this type of mechanism is actively discussed.~

 We have demonstrated previously~\cite{Boyarsky:11a} that \emph{there exists another
 (new) mechanism for the transfer of magnetic helicity} from short wavelengths to
 long wavelengths. This mechanism is \emph{not} related to
 magneto-hydro\-dynamical turbulence, but has its root in the \textit{chiral
 anomaly}. We have then argued that, because of their coupling to the effective
 axion field (which describes chiral asymmetries), helical magnetic fields may
 survive down to rather low temperatures of a few MeV, even though chirality
 flipping reactions (caused by mass terms for the charged particles) are in
 thermal equilibrium. It has also been argued in~\cite{Boyarsky:12a} that, as
 a consequence of the parity-violating nature of weak interactions and the
 chiral anomaly, quantum corrections in states of finite lepton- or baryon
 number density may give rise to an asymmetry between left-handed and
 right-handed particles. A time-dependent effective axion field is then
 generated, which, in turn, excites a magnetic field
 \cite{Frohlich:2000en}.  One concludes that a homogeneous, isotropic
 stationary state of the Standard-Model plasma might be
 unstable. Thus, one has to study \textit{dynamical} features of this system,
 in order to identify its true equilibrium state.

Large-scale electromagnetic fields in the plasma excite macroscopic flows of 
matter consisting of charged particles. The dynamics of low-energy modes of the plasma
can be described with the help of the equations of relativistic magneto-hydrodynamics
(RMHD). One might expect that the hydrodynamical
description of a magnetized relativistic plasma does not differ in an essential way from the one of its
non-relativistic cousin, the only important difference being a different relation
between energy density, $\rho$, and pressure, $p$; (~for reviews see, e.g.,
\cite{Kandus:10,Subramanian:2009fu,Banerjee:04}).  However, the
chiral anomaly actually leads to an important  modification of the Navier-Stokes equation; see
e.g.~\cite{Son:2009tf,Kharzeev:2011vv,Kharzeev:11a}. Furthermore, helical 
electromagnetic fields affect the evolution of the difference of
chemical potentials of left- and right-handed particles, which becomes (space- and) 
time-dependent~\cite{Boyarsky:11a}. As a result, not only the Navier-Stokes equations,
but also the connection between the electric current density and
the magnetic field are modified in accordance with
the chiral anomaly, and an equation of motion for an effective axion field
must be added. For a spatially homogeneous state, the time derivative of this 
axion field is nothing but the chiral chemical potential, (i.e., the difference of chemical 
potentials for left- and right-handed particles).

A \emph{magneto-hydrodynamical} description of a chiral relativistic plasma of
charged \emph{massive} particles, as studied in this paper, provides an
approximate description of the Standard-Model plasma at
high temperatures and positive matter density; (additional effects not related to the
chiral anomaly must, however, be taken into account, too). We expect that our 
magneto-hydrodynamical description of the plasma is a useful and reliable tool to 
determine properties of (local) thermal equilibrium of the plasma at non-vanishing 
lepton- and baryon densities.

Apart from applications in studies of the
Early Universe, our magneto-hydrodynamical approach may lead to a reasonably accurate
description of astrophysical systems, such as
relativistic jets, and of \emph{heavy ion collisions}~\cite{Kharzeev:2011vv}.  
Moreover, it may change our current views of
the origin and evolution of primordial magnetic fields and their
relationship to the presently observed large-scale cosmic magnetic
fields~\cite{Neronov:10a,Tavecchio:10,Dolag:10}. Our results and the effect reported
in~\cite{Boyarsky:12a} could affect our current understanding of cold but very 
dense systems, such as neutron stars. Last but not least, our magneto-hydrodynamical approach 
may be useful in the analysis of models of beyond-the-Standard-Model particle physics, 
such as the $\nu$MSM~\cite{Asaka:05a,Asaka:05b,Boyarsky:09a}.  

We stress that the effects studied in this paper do \emph{not} involve any
``new physics''; they can be described within the Standard Model. 
The axion field introduced below might not be a fundamental field but, rather,
an emergent one that appears in the description of certain states of the
Standard-Model plasma. If instead a fundamental axion \emph{does exist}, then
the high temperature states of Standard Model plasma augmented by this new
field (as in~\cite{Frohlich:2002fg,Frohlich:2000en}) can also be described by
the chiral RMHD and the effects, discussed in this paper could emerge in a
wider class of states.

\subsection{Main goals of the paper}
\label{sec:prev-work}

The physical system of primary interest in this paper is a relativistic plasma
of very light charged particles at a temperature $T$ large enough that the
masses of the lightest charged particles, the electrons and positrons, can be
neglected, (i.e., set to $0$). For simplicity, one may consider a plasma consisting
only of electrons and positrons, with a charge-neutralizing background of
protons (and neutrons). But our analysis readily extends to more general systems.

Let $J^\mu_5(x)$ denote the usual axial vector current density, where
$x=(t,\vec x)$ are coordinates of a space-time point; $\vec E(x)$ and $\vec
B(x)$ denote the electric field and magnetic induction,
respectively. (In our notations, we will not distinguish the quantized electromagnetic field from its expectation value in physically interesting states.) 
Furthemore, $A(x) = (A_0(x)\equiv \phi(x), \vec A(x))$ is the
electromagnetic gauge field, with $\phi$ the electrostatic potential and $\vec
A$ the vector potential. The chiral anomaly says that $J^\mu_5$ is \emph{not}
a conserved current; more precisely that
\begin{equation}
    \label{eq:6}
    \partial_\mu J^\mu_5(x) = \frac{2\alpha}{\pi} \vec E\cdot \vec B(x)\;,
  \end{equation} 
where $\alpha$ is the fine structure constant.
Introducing the current density
\begin{equation}
  \label{eq:136}
  \tilde J^\mu_5 \equiv  J^\mu_5- \frac{\alpha}{2\pi}
  \epsilon^{\mu\nu\rho\sigma} A_\nu F_{\rho\sigma},
\end{equation}
where $F_{\rho\sigma}$ are the components of the electromagnetic field tensor,
we find that 
\begin{equation}
  \label{eq:138}
  \partial_\mu \tilde J^\mu_5 = 0,
\end{equation}
i.e. that $\tilde J^\mu_5$ is a \emph{conserved current}
density. Unfortunately, it is \emph{not} invariant under electromagnetic gauge
transformations, $A_\mu \to A_\mu - \partial_\mu \eta$, where $\eta$ is
an arbitrary function on space-time. However, the \emph{charge}
\begin{equation}
  \label{eq:196}
\tilde Q_{5} \equiv \int\limits_{t=\const} \hskip -1.5ex d^3\vec
x\:\tilde J^0_5 (\vec x,t)
\end{equation}
is not only \emph{conserved} but also \emph{gauge-invariant}. In order to
describe thermal equilibrium states of the plasma, we are advised to
introduce an ``\emph{axial chemical potential}'', $\mu_5$, conjugate to
$\tilde Q_5$. 
The chemical potential conjugate to the conserved total electric charge is tuned in
such a way that the system is \emph{neutral}; (i.e., the charge density of
electrons and positrons in the system is canceled, in average, by the one of the
heavy charged hadrons, in particular the protons).

We introduce the pseudo-scalar density $$\rho_5(\vec x,t) \equiv \langle
J^0_5(\vec x,t)\rangle_{T,\mu_5}$$ and its spatial average
$$q_5 \equiv \overline {\rho_5(\vec x,t)},$$
where we denote by $\langle (\cdot) \rangle_{T,\mu_5}$ the equilibrium state of
the system at temperature $T$ and axial chemical potential $\mu_5$, and by
$\overline{(\cdot)}$ we indicate spatial averaging. As one would expect, there is a
response equation relating the average axial charge density, $q_5$, to the axial
chemical potential $\mu_5$ of the form
\begin{equation}
  \label{eq:197}
  q_5(T,\mu_5) \approx \mu_5 \pfrac{q_5}{\mu_5}\biggr|_{\mu_{5}=0} \approx \frac{\mu_5}6 T^2\;, 
\end{equation}
with $q_{5}(T, \mu_{5}=0) = 0$. A ``derivation'' of this relation can be found
in cosmology text books.\footnote{It is assumed that in processes such as
  $e^{+} + e^{-} \leftrightarrow \gamma + \gamma$ and $e^\pm + \gamma
  \leftrightarrow e^\pm + \gamma$ the rates of direct and inverse reactions
  are identical, see, e.g.,~\cite{Kolb-Turner}, Sect. 3.4 and Chapter 5.}

Using the chiral anomaly, we see that Eq.~(\ref{eq:197}) implies the equation
\begin{equation}
  \label{eq:198}
  \dot{\mu}_{5} = \frac{6}{T^{2}}\dot{q}_{5}(T, \mu_{5}) = \frac{12
    \alpha}{\pi T^{2}} \overline{\strut \vec E \cdot \vec B}
\end{equation}
a relation derived and explained in~\cite{Boyarsky:11a}. (The dot indicates a
derivative w.r. to time, and we neglect terms $\sim \dot{T}$!) It has been
shown in \cite{Frohlich:2000en,Alekseev:98a} that, in the presence of a
non-vanishing magnetic induction $\vec B (x)$, a non-zero axial chemical
potential, $\mu_{5} \neq 0$, induces an electric current density given 
by $\frac{\alpha}{\pi}\mu_{5} \vec B(x)$, so that the
complete expression for the electric current density, $\vec j$, of the plasma is given by
\begin{equation}
  \label{eq:199}
  \vec{j}(x) = \langle \vec J(x) \rangle_{T, \mu_{5}} = \frac{\alpha}{\pi}\mu_{5}\vec B(x) + \sigma \vec E (x) \;,
\end{equation}
where $\sigma$ is the Ohmic conductivity.\footnote{For simplicity, the state
  of the plasma is assumed to be isotropic and homogeneous.} In
\cite{Frohlich:2000en,Alekseev:98a}, the term $\frac{\alpha }{\pi }\mu_{5}\vec
B$ on the right side of~(\ref{eq:199}) has been derived from the chiral
anomaly using current algebra; (see \cite{Treiman:85}).

Inserting~(\ref{eq:199}) in Amp\`ere's law and applying Faradey's induction
law, one readily finds the equation\footnote{The standard approximation made
  in MHD is to neglect Maxwell's displacement current and hence the second
  derivative, $\frac{\partial^{2}\vec B}{\partial t^{2}}$, on the right side
  of ~(\ref{eq:200}); see, e.g., \cite{Dorch:2007}.}
\begin{equation}
  \label{eq:200}
  \frac{\partial \vec B}{\partial t} = \frac{1}{\sigma}\square \vec B + \frac{\alpha}{\pi}\frac{\mu_{5}}{\sigma}\vec{\nabla}\wedge \vec B \;. 
\end{equation}
As shown in \cite{Frohlich:2000en,Joyce:97}, the term proportional to
$\mu_{5}$ in Eq.~(\ref{eq:200}) is responsible for an \emph{instability} of
the solutions of~(\ref{eq:200}), namely an exponential growth of the magnetic
induction $\vec B$, which leads to a non-vanishing density $\vec E \cdot \vec
B$, (i.e., to a non-vanishing helicity of the electromagnetic field). By
eq.~(\ref{eq:198}), this, in turn, leads to a temporal variation of the
\emph{axial} chemical potential $\mu_{5}$ and, hence, to a change in time of
the pseudo-scalar density $q_{5}$ that describes the asymmetry between the
charge densities of left-handed and right-handed charged particles. (We note
that the electric field is damped by Ohmic losses, which affects the time dependence
of $\vec{E} \cdot \vec{B}$.) 

Equation~(\ref{eq:198}) implies that, in general, $\mu_{5}$ depends on time.
The anomaly equation~(\ref{eq:6}) shows that the chiral density $\rho_{5}
(\vec x , t)$ usually not only depends on time, but also on \emph{space},
$\vec x$. It is therefore necessary to generalize relation~(\ref{eq:197}) to
one where $\mu_{5}$ \emph{not only depends on $t$, but also on space $\vec
  x$}; to then derive the correct form of the Maxwell equations in the presence of a
space- and time-dependent axial chemical potential, $\mu_{5}(\vec x, t)$, and to
derive the equations of motion for this field.  \textit{To find such a generalization
is one of the main goals of this paper}.  We will also study how to couple our
system of equations to the equations of motion of a relativistic fluid, thereby 
deriving the correct equations of relativistic
magneto-hydrodynamics (RMHD). In Sect.~\ref{sec:one-mode-solution} we discuss
some simple solutions of these equations and sketch possible applications to plasma
physics, astrophysics and cosmology.

As was pointed out already in \cite{Frohlich:2000en}, electrodynamics
coupled to an axion field, $\theta_5$, can be used to describe the dynamics of a 
homogeneous system (with $\theta_5 = \theta_5(t)$ independent of $\vec{x}$) 
in the presence of an axial chemical potential
$\mu_5(t) \equiv \dtheta_5(t)$ satisfying Eq.~\eqref{eq:198}.
In this work, we show that, for an
\textit{inhomogeneous} chemical potential $\mu_5(\vec x,t)$ (defined below,
Sect.~\ref{sec:LTE_DM}), equations analogous to Eqs.~\eqref{eq:198}
and~\eqref{eq:200} can be written as \emph{local} field equations for the electromagnetic field 
and for an effective axion field $\theta_{5}(x)$, with $\dtheta_5(\vec x,t)=\mu_5(\vec x,t) $, (rather than 
in terms of the axial chemical potential $\mu_5(\vec{x},t)$), as already pointed 
out in \cite{Frohlich:2000en}.
The axion field $\theta_5(\vec x,t)$ interacts with the electromagnetic
field in the form of a term $\theta_{5}\vec{E}\cdot \vec{B}$ in the Lagrangian density,
but its dynamics is not necessarily described by a
relativistic wave equation, but might be governed by an inhomogeneous diffusion equation.

So far, the effective axion field $\theta_5$ has been treated as a
\textit{classical} field.  The role played be the axial chemical potential
$\mu_5(\vec x,t)$ in the description of inhomogeneous \textit{local} thermal
equilibrium states suggests that $\mu_5$, and hence the axion field $\theta_5$
does not describe hamiltonian degrees of freedom that have to be quantized;
but that they should be understood as $c$-number fields labeling a family of
generally inhomogeneous non-stationary states of the quantum-mechanical plasma
-- as it is the case of the local \emph{temperature} (\emph{field}) $T(x)$.

One could also consider a system with an axion field \textit{does} describing
dynamical (hamiltonian) degrees of freedom that must be quantized. In order
to shed new light on this important issue, we show, in Appendix~\ref{sec:5d},
how an effective axion field emerges in theories with extra dimensions.  In
particular, we consider a model of (quantum) electrodynamics on a slab in
\textit{five-dimensional} Minkowski space in the presence of a 5D Chern-Simons
term.  In this theory, the axion appears as a component of the electromagnetic
gauge field (in the fifth direction transversal to the boundaries of the
slab), and the state function $T(x)$ has an interpretation as the inverse
width of the slab and is therefore related to the geometry of five-dimensional
space-time; (see also \cite{Chamseddine:12}). Comparing this 5D formulation
with the analysis presented in Sects.~\ref{sec:LTE_DM}
and~\ref{sec:dynamics-chiral-asymmetry}, one may wonder whether, in the end,
space-time geometry might merely encode properties of the \emph{state} of
quantum-mechanical degrees of freedom, rather than being described by
\textit{dynamical} quantum fields (in particular, a quantized metric
field). These fundamental questions deserve further study with a view towards
applications in cosmology.

\section{Expression for the electric current density in the presence of an inhomogeneous axial chemical potential}
\label{sec:LTE_DM}

\subsection{States describing local thermal equilibrium}
  \label{sec:LTE_Q5}

  In order to account for the local nature of the asymmetry between left-handed
  and right-handed particles, we propose to introduce a space- and
  time-dependent axial chemical potential, $\mu_5(x)$, and generalize
  Eq.~(\ref{eq:197}) to a local relation between the pseudo-scalar density $
  \rho_5$ and $\mu_5$, valid in the regime where $\mu_5 \ll T$. We assume that
  \textit{local thermal equilibrium (LTE)} is reached at length scales \textit{small} as
  compared to the scale of spatial variations of $\mu_5(x)$, i.e.,
  $l_\text{LTE} \ll \mu_5(x)/|\nabla \mu_5(x)|$), and that the scale of
  variation of electromagnetic fields and matter flows is much larger than
  $l_\text{LTE}$, too.  

  A quantum-mechanical system in local thermal equilibrium can be
  described similarly as one in perfect equilibrium; (see, e.g.,
  \cite{zubarev1974nonequilibrium}). Its state is tentatively described by the
  density matrix
\begin{equation}
  \label{eq:9}
  \varrho_\text{\sc lte} \equiv \mathcal{Z}^{-1} \exp\left\{-\frac{\CH-
      \int_{y}
      \mu_5(y) \tilde{J}_{5}^{0}(y)
    }T\right\}
\end{equation}
Here $\CH$ is the Hamiltonian of the system, and $\mathcal{Z}$ is the
partition function (chosen such that $\ \tr(\varrho_\text{\sc lte})\equiv 1
$); $\tilde J^{0}_5$ is as in Eqs.~\eqref{eq:136} and~\eqref{eq:196}, and $T$
is the temperature (here assumed to be constant throughout space-time). 

The integration in the exponent on the right side of \eqref{eq:9} extends over
a spatial hyperplane at fixed time. Expression \eqref{eq:9} may be generalized
as follows: Let ($u^{\mu}(x)$) be the hydrodynamical 4-velocity field of the
plasma; (the condition that $u^{0}(x) \equiv 1$, $u^{i}(x) \equiv 0$, for
$i=1,2,3$ means that we are working in a co-moving coordinate frame). Let
($T_{\mu \nu}(x)$) denote the components of the energy-momentum tensor of the
system, and let $\beta(x) \equiv (k_{B}T(x))^{-1}$ be a (possibly slowly
space- and time-dependent) inverse-temperature field. Let $\Sigma$ be some
space-like hypersurface, and let $d \sigma^{\mu}(x)$ denote the (\emph{dual}
of the) \emph{surface element} at a point $x \in \Sigma$. The covariant form
of the expression \eqref{eq:9} is given by (see e.g.\ the
book~\cite[Chap. 24]{zubarev1974nonequilibrium}):
\begin{equation}
  \label{eq:10}
  \rho_\text{\sc lte}=\mathcal{Z}_{\Sigma}^{-1} \exp\left\{-\int \limits_{\Sigma}^{}\left[u^{\mu}(x)T_{\mu \nu}(x) - \mu_5(x) \tilde J_{5\nu}(x)\right]\beta (x) d \sigma^{\nu}(x)
  \right\}
\end{equation}
An equivalent expression is given in Sect.~\ref{sec:axion-degree-freedom},
Eq.~(\ref{eq:10'}). Expectation values of operators in the state introduced in 
\eqref{eq:10} are denoted by $\langle (\cdot) \rangle_{\text{LTE}}$. We recall from Eq.~(\ref{eq:136}) that
$$
\tilde J_5^{\mu}=J_5^{\mu} - \frac{\alpha}{2\pi}\varepsilon^{\mu \nu \rho
  \sigma} A_\nu F_{\rho \sigma},
$$ 
which, by Eq.~(\ref{eq:6}) (chiral anomaly), is conserved but \emph{not}
gauge-invariant. Under a gauge transformation, $A_{\nu} \to A_{\nu}
- \partial_{\nu}\eta$, where $\eta$ is an arbitrary (smooth) function on
space-time, the term
$$
\int \limits_{\Sigma}^{} \beta (x) \mu_{5}(x)\tilde J_{5
  \nu}(x)d\sigma^{\nu}(x) = \int \limits_{\Sigma}^{} \beta (x)\mu_{5}(x)\tilde
J_{5}^{\nu}(x)d\sigma_{\nu}(x)
$$
changes by
$$
  -\frac{\alpha}{2 \pi}\int_{\Sigma} \varepsilon^{\nu \mu \rho \lambda}\partial_{\mu}\left( \beta (x)\mu_{5}(x) \right)F_{\rho \lambda}(x)\eta (x) d\sigma_{\nu}(x),
$$
as follows by integration by parts; using that $\partial_{[ \mu}F_{\rho
  \lambda ]} \equiv 0$. Requiring that expression~(\ref{eq:10}) be
gauge-invariant (i.e., independent of the gauge function $\eta $, for
arbitrary $\eta$), we find the constraint
\begin{equation}
  \label{eq:11}
   \varepsilon^{\mu \nu \rho \sigma}\partial_{\nu}\left( \beta (x) \mu_{5}(x)\right) F_{\rho \sigma}                          [d\sigma^{\mu}(x)]_{\Sigma}^{} \equiv 0,
\end{equation}
or, in the language of differential forms,
\begin{equation}
  \label{eq:1111}
   d\left( \beta \mu_{5}\right) \wedge F\vert _{\Sigma}^{} \equiv 0.
\end{equation}
If $\beta (x) \equiv \beta $ is constant and $\Sigma$ is a hyperplane at fixed time then (\ref{eq:11}) implies that
\begin{equation}
  \label{eq:12}
   \vec B(x) \cdot \vec \nabla \mu_{5}(x) \equiv 0,
\end{equation}
i.e., gauge invariance implies that the gradient of the axial chemical
potential is orthogonal to the magnetic induction. Condition~(\ref{eq:12}) has
a simple physical interpretation discussed in Sect.~\ref{sec:grad-mu5}, below.

Let $J^{\mu}(x)$ denote the electric (vector) current density, and let
$J^{\mu}_{5}$ be as in Sect.~\ref{sec:prev-work}; (see Eq.~(\ref{eq:6})). We
set
\begin{equation}
  \label{eq:13}
  j^{\mu}(x) = \langle J^{\mu}(x)\rangle_\text{\sc lte}, \quad j_{5}^{\mu}(x) = \langle J_{5}^{\mu}(x)\rangle_\text{\sc lte},
\end{equation}
for $\mu = 0,1,2,3$. In \cite{Frohlich:2000en,Frohlich:2002fg}, 
the equal-time commutation relations
between $J^{0}$ and $J^{0}_{5}$,
\begin{equation}
  \label{eq:14}
  [ J^{0}_{5}(\vec y , t), J^{0}(\vec x , t)] = \frac{ \alpha}\pi\vec B (\vec y , t) \cdot \vec \nabla_{y} \delta (\vec x - \vec y),
\end{equation}
see~\cite{Treiman:85}, have been used to derive expression (\ref{eq:199}) for the $\vec{B}$- dependence of the electric current density $\vec{j}$:
\begin{equation}
  \label{eq:15}
  \vec \jmath\, (x) = \frac{\alpha}{\pi}\mu_{5}\vec B(x),
\end{equation}
assuming that $\mu_{5}$ does not depend on $\vec{x}$ and that $\sigma \vec E \equiv 0$. We sketch how (\ref{eq:15}) is derived from (\ref{eq:14}) and then propose a generalization of (\ref{eq:15}) where $\mu_{5}$ may depend on space and time.

The conservation of the vector current density $J^{\mu}(x)$ implies that there exists a "current vector potential", $\vec \Phi (x)$, with
\begin{equation}
  \label{eq:16}
  J^{0}(x) = -\vec \nabla \cdot \vec \Phi (x), \quad \vec J(x) = \partial_t \vec \Phi (x).
\end{equation}
These expressions are invariant under the transformations $\vec \Phi \to \vec \Phi + \vec \nabla \wedge \vec X$, for an arbitrary vector field $\vec X$ independent of time. The commutation relations of the field $\vec \Phi$ with the axial charge density $J_{5}^{0}$ are then given by
\begin{equation}
  \label{eq:17}
  [J^{0}_{5}(\vec y , t),\vec\Phi (\vec x, t)] = \frac{\alpha}{\pi} \vec B (\vec y,t)\delta(\vec x- \vec y) + \vec \nabla_{x} \wedge \vec \Pi (\vec x , \vec y, t),
\end{equation}
where the second term on the right side of (\ref{eq:17}) is an "integration constant". Let $\vec \pi$ denote the expectation value of $\vec \Pi$ in the state $\langle ( \cdot ) \rangle_\text{\sc lte}$ of the plasma. If this state is homogeneous (in particular, $\langle \vec B \rangle_\text{\sc lte} = 0$) then $\vec \pi (\vec x , \vec y, t) \equiv \vec \pi (\vec x - \vec y , t)$ must be space-translation invariant. It is plausible that $\vec \pi$ is independent of $\vec B$, so that $\vec \pi$ is space-translation invariant even if $\langle \vec B \rangle_\text{\sc lte} \neq 0 $. In the following, we will write $\vec B $ for $\langle \vec B \rangle_\text{\sc lte}$, as above, in order to shorten our notation. 
Taking the expectation value of both sides of (\ref{eq:17}) in the state 
$$
\langle (\cdot )\rangle \equiv \langle (\cdot )\rangle_\text{\sc lte}
$$ 
of the plasma and integrating over $\vec y$, we find that
\begin{equation}
  \label{eq:18}
  \langle [Q_{5}, \vec \Phi (x)]\rangle = \frac{ \alpha}{\pi} \vec B (x),
\end{equation}
because the expectation of the second term on the right side of (\ref{eq:17}) integrates to $0$ if $\vec \pi = \langle \vec \Pi \rangle $ is translation-invariant. 

Since $\vec \Phi$ commutes with the electric charge operator $Q$, eq. (\ref{eq:18}) implies that
$$
\langle [Q_{L / R},\vec \Phi (x)] \rangle = \pm \frac{\alpha}{2\pi}\vec B (x),
$$
where $Q_{L/R} = \frac{1}{2}\left( Q \pm Q_{5}\right)$. 

Next, we observe that 
\begin{equation}
  \label{eq:19}
  \vec j (x) = \langle \dot{\vec{\Phi}} (x)\rangle = \text{i tr} \left( \rho_\text{\sc lte}[\mathcal{H} , \vec \Phi (x)]\right),
\end{equation}
where $\rho_\text{\sc lte}$ is given by (\ref{eq:9}), with $\mu_{5}(x) \equiv \mu_{5}$ independent of $x$. 

Formally, the right side of (\ref{eq:19}) appears to vanish for a constant $\mu_{5}$, because $\tilde \jmath^{\mu}_{5}$ is a conserved current.
However, the field $\vec \Phi$ is so singular in the infrared that the formal calculation is deceptive. The right side of (\ref{eq:19}) must be regularized in the infrared by adding a mass term to the Hamiltonian, 
$\mathcal{H} \to \mathcal{H}_{mass}$. Then $\tilde \jmath_{5}^{\mu}$ is \emph{not} conserved, anymore, and $[\mathcal{H}_{mass}, Q_{5}] \neq 0$. However $\rho_\text{\sc lte}$ continues to commute with $\mathcal{H}_{mass} - \mu_{5}\int \tilde \jmath_{5}^{0}(\vec y)d^{3}y$! Thus
\begin{equation}
  \label{eq:20}
  \begin{aligned}
    \vec j (x) = &\langle \dot{\vec{\Phi}} (x)\rangle = \text{i tr} \left(
      \rho_\text{\sc lte} \left[ \mathcal{H}_{mass} - \mu_{5}\int \tilde
        \jmath^{0}_{5} (\vec y , t) d^{3}y + \mu_{5}\int \tilde
        \jmath_{5}^{0}(\vec y , t) d^{3}y, \vec \Phi (\vec x ,
        t)\right]\right) \\=&\mu_{5}\langle [Q_{5}, \vec \Phi (x)]\rangle =
    \frac{ \alpha }{\pi}\mu_{5}\vec B (x),
  \end{aligned}
\end{equation}
by Eq.~(\ref{eq:18}).

Next, we consider the general situation where $\mu_{5}$ may depend on time $and$ space. Let $\vec{\pi}(\vec{x},\vec{y},t) := \langle\vec \Pi (\vec x , \vec y , t)\rangle$, where $\vec{\Pi}$ is as in Eq.~(\ref{eq:17}). 
Then Eq.~(\ref{eq:20}) generalizes to
\begin{equation}
  \label{eq:21}
  \vec \jmath\, (x) = \frac{ \alpha }{\pi}\mu_{5}(x)\vec B (x) + \vec \nabla_{x}\wedge \int \vec \pi (\vec x,\vec y, t) \mu_{5}(\vec y , t) d^{3}y,
\end{equation}
Note that the current $\vec j$ in (\ref{eq:21}) is conserved as a consequence
of Eq.~(\ref{eq:12}); (gauge-invariance of $\rho_\text{\sc lte}$). If $\mu_5$
is independent of spatial coordinates and the distribution $\vec{\pi}$ is translation-invariant, the second term in Eq.~\eqref{eq:21}
vanishes, and this equation reduces to \eqref{eq:20}.

\subsection{Axial chemical potential and axion field}
\label{sec:axion-degree-freedom}

The second term in the expression~\eqref{eq:21} is non-local and the distribution
$\vec \pi (\vec x, \vec y, t) $ has not been defined, yet. We attempt to find
a local expression for the four-current density $j^{\mu}$. This current density should be
proportional to $(F_{\mu\nu})$ and linear in $(\mu_5)$; and it should be conserved, (i.e., satisfy 
the continuity equation). By introducting a pseudo-scalar field,
$\theta_5(x)$, indeed an \emph{axion field}, related to the axial chemical
potential $\mu_5(x)$, we arrive at the following ansatz for $j^{\mu}$:
\begin{equation}
\label{eq:22}
j^\mu_\axion \equiv \frac{\alpha}{2\pi}\epsilon^{\mu\nu\rho\sigma}(\partial_\nu
\theta_5)(x)
F_{\rho\sigma}(x),
\end{equation}
with $\dot{\theta}_{5}$ proportional to $\mu_{5}$ in co-moving coordinates.
The current $j^\mu_\axion$ defined by Eq.~\eqref{eq:22} is automatically
conserved; ($j_{\text{axion}}^{\mu}$ is dual to the 3-form
$\frac{\alpha}{2\pi}d \theta_5 \wedge F$, which is \emph{closed}, because
$d^2=0=dF$).  For $j_{\text{axion}}^{\mu}$ to transform as a \emph{vector}
under parity and time reversal, $\theta_5$ must be a pseudo-scalar
field. 

To define its relation to $\mu_5$ let us 
separate the 0-components from the
spatial components:
\begin{align}
\label{eq:23}
j^0_\axion = \dfrac{\alpha}\pi\vec \nabla \theta_5\cdot \vec B, \qquad
\vec\jmath_\axion = \dfrac{\alpha}\pi\Bigl(\dot \theta_5 \vec B + \nabla
\theta_5 \wedge \vec E\Bigr).
\end{align}
Comparing these equations with Eq.~(\ref{eq:15}) (see also
Eq.~(\ref{eq:199})), we argue that $ \mu_5(x)=\dot\theta_5(x) $. The
identification of the axial chemical potential with an axion field
($\dot\theta_5$) has first been proposed
in~\cite{Frohlich:2000en,Frohlich:2002fg}; (later it also appeared in the
context of the quark-gluon plasma in
Refs.~\cite{Kalaydzhyan:2012ut,Huang:2013iia,Zhitnitsky:2012im,Kharzeev:2007tn,Landsteiner:2012kd,Kharzeev:2011rw},
with the pseudo-scalar field being identified with a variable QCD theta
angle).

We wish to explain the relation between an inhomogeneous $\mu_5(x)$ and
$\theta_5(x)$. Eq.~\eqref{eq:23} holds in any coordinate system, if we define
$E_i = F_{0i}$ and $B_i = \frac12 \epsilon_{ijk}F^{jk}$. However, in plasma physics, it
is convenient to express $F_{\mu\nu}$ in terms of electric and magnetic fields in
co-moving coordinates:
\begin{equation}
  \label{eq:27}
  F_{\mu\nu} = \epsilon_{\mu\nu\lambda\rho}u^\lambda {\cal B}^\rho + (u_\mu {\cal E}_\nu -
  u_\nu {\cal E}_\mu),
\end{equation}
where ${\cal E}_\mu,{\cal B}_\mu$ are 4-vectors of electric and magnetic
fields orthogonal to the 4-velocity field $u^\mu$, (i.e., ${\cal E}_\mu u^\mu = {\cal B}_\mu
u^\mu =0$). For details see e.g.\ the
review~\cite{Subramanian:09}). Taking $F_{\mu\nu}$ in the form~\eqref{eq:27}
and plugging it into the Eq.~\eqref{eq:22}, we find that
\begin{equation}
  \label{eq:80}
  \vec j_\axion = \frac\alpha\pi\biggl[ \bigl(\dtheta_5 + \vec v \cdot\vec
  \nabla \theta_5\bigr)\vec {\cal B}(x) + \bigl(\vec\nabla \theta_5 + \vec v\, \dtheta_5\bigr)\wedge
  \vec {\cal E}(x)\biggr]+\CO(v^2)
\end{equation}
(where $\vec {\cal E}, \vec {\cal B}$ are the spatial components of the
4-vectors ${\cal E}_\mu, {\cal B}_\mu$. In what follows, we denote these
fields again by $\vec E, \vec B$).  Comparing Eq.~\eqref{eq:80}
with~\eqref{eq:21} gives us the desired relation between $\mu_5$ and
$\theta_5$
\begin{equation}
\label{eq:24}
\mu_5(\vec x,t) \equiv u^\mu \partial_\mu \theta_5(\vec x,t).
\end{equation}
We note that if the relation $\mu_5(t) = \dtheta_5(t)$ holds in co-moving
coordinates then Eq.~\eqref{eq:24} follows by transformation to the
laboratory frame. 

Eqs.~\eqref{eq:24} and \eqref{eq:80}, \eqref{eq:23} provide a local expression for
the electric four-current density $j^{\mu}$ involving the axion field $\theta_5$. Our expression
for $j^{\mu}$ explains why the electric current density $\vec{j}$ in ~\eqref{eq:21} is non-local 
in the axial chemical potential $\mu_5$. The current density $\vec{j}$ is \emph{odd} under
 \emph{time reversal}, $\mathcal{T}$.  The combination $\mu_5 \vec
B$ is $\mathcal T$-odd, too, because $\vec B$ is $\mathcal T$-odd and $\mu_5$ is
$\mathcal T$-even.  The electric field ($\vec E$) is
$\mathcal T$-even, hence a term proportional to (spatial derivatives of)
$\mu_5$ and $\vec E$ would have the wrong transformation
properties under time reversal. However, when introducing the axion
field $\theta_5$, it is easy to write a local combination of (spatial derivatives of) $\theta_{5}$ 
and $\vec{E}$ that transforms correctly under time reversal, namely
$\vec{\nabla} \theta_5 \cdot \vec{ E}$, which is odd under $\mathcal{T}$, because the axion field is 
$\mathcal{ T}$-odd, (as the
expression in~\eqref{eq:24} shows). Clearly, $\vec\nabla \theta_5 $ is a non-local 
functional of $\mu_5$.  (Thus, introducing an axion field enables one to find a
 \emph{local} expression for the electric current, instead of the
non-local one on the right hand side of Eq.~\eqref{eq:21}, and the second
equation in (\ref{eq:23}) is the correct generalization of Eq. \eqref{eq:20}
to a spatially inhomogeneous plasma.

An alternative route towards understanding the origin of the axion field $\theta_{5}$  is
the following one, (see ~\cite{Frohlich:2000en,Frohlich:2002fg}): We consider a system of massless
fermions coupled to the electromagnetic field in a state corresponding to a non-vanishing axial
chemical potential $\mu_{5}$. The action of this system is given by
\begin{equation}
  \label{eq:116}
  S[A,\psi] = 
  \int d^4x\, \left[-\frac14 F_{\mu\nu}^2 +
    \bar\psi \Bigl(\slashed{\partial} + e\slashed{A}\Bigr)\psi + 
    \mu_5 (\bar\psi_L^\dagger \psi_L - \bar\psi^\dagger_R\psi_R)\right] 
\end{equation}
One can remove the term proportional to $\mu_{5}$ by performing a \emph{``chiral
  redefinition''} of the fermion fields
\begin{equation}
  \label{eq:117}
  \psi \to e^{i \theta_5(t)\gamma_5}\chi
\end{equation}
where $\dot\theta_5 = \mu_5$; (see ~\eqref{eq:24}). This transformation
yields a Fujikawa Jacobian~\cite{Fujikawa:1979ay} 
\begin{equation}
  \label{eq:134}
  \mathcal{J}[A_\mu,\theta_5] = \exp \left(i\,\frac{\alpha}{4\pi} {\theta_5}\epsilon^{\mu\nu\lambda\rho}F_{\mu\nu}
 F_{\lambda\rho} \right)
\end{equation}
in the functional integral, yielding the modified action
\begin{equation}
  \label{eq:118}
  S[A,\chi,\theta_5] = \int d^4x\, \left(-\frac14 F_{\mu\nu}^2 +\,\frac{\alpha}{4\pi} {\theta_5} \epsilon^{\mu\nu\lambda\rho}F_{\mu\nu}
    F_{\lambda\rho} +
    \bar\chi \Bigl(\slashed{\partial} + \slashed{A}\Bigr)\chi\right).
\end{equation}
We see that if the axial chemical potential is spatially homogeneous then
it can be re-expressed in terms of a pseudo-scalar field, $\theta_5$, that
has an axion-like interaction with the electromagnetic field. 

Eq.~(\ref{eq:24}) enables us to present the general expression for the density
matrix describing \emph{local thermal equilibrium} of an inhomogeneous plasma
in an arbitrary coordinate system.
\begin{equation}
\label{eq:10'}
\rho_\text{\sc lte}=\mathcal{Z}_{\Sigma}^{-1} \exp\left\{-\int \limits_{\Sigma}^{}\Bigl [u^{\mu}(x)\bigl(T_{\mu \nu}(x) - \partial_{\mu}\theta_5(x) \tilde{J}_{5\nu}(x)\bigr)\Bigr]\beta (x) d \sigma^{\nu}(x)\right\},
\end{equation}
where $u^{\mu}$ is the 4-velocity field of the plasma and the expression
$u^\mu \partial_\mu \theta_5$ plays the role of a local axial chemical potential
$\mu_5(x)$. Expression ~\eqref{eq:10'} generalizes \eqref{eq:10}
to an arbitrary coordinate system.

In Section~\ref{sec:dynamics-chiral-asymmetry} we will search for the correct
dynamics of the axion field. But, before addressing this problem, we propose to
clarify the significance of the constraint~\eqref{eq:12}.

\subsection{Constraints on $\theta_5$} 
\label{sec:grad-mu5}
In this subsection, we tentatively interpret $\theta_5$ as a $classical$ field, more precisely as a generalized thermodynamic parameter that labels states describing \emph{local} thermal equilibrium with left-right (chiral) asymmetry. As discussed in Sects.~\ref{sec:LTE_Q5} and \ref{sec:axion-degree-freedom}, the requirement that the density matrix $\rho_\text{\sc lte}$, supposed to describe local thermal equilibrium (see (\ref{eq:10}), (\ref{eq:10'})) be gauge-invariant yields the constraint 

\begin{equation}
d(\beta\mu_5)\wedge F\big|_{\Sigma}^{}\equiv 0,
\label{eq:32}
\end{equation}
with 
\begin{equation}
  \mu_5=u^{\mu}\partial_\mu\theta_5,
\label{eq:33}
\end{equation}
where $u^{\mu}(x)$ is the 4-velocity field of the plasma appearing in expressions (\ref{eq:10})
and (\ref{eq:10'}) for $\rho_\text{\sc lte}$. Eq.~(\ref{eq:32}) is identical to
(\ref{eq:11})--(\ref{eq:1111}), and Eq.~(\ref{eq:33}) follows from comparing
expressions (\ref{eq:10}) (Sect.~\ref{sec:LTE_Q5}) and (\ref{eq:10'})
(Sect.~\ref{sec:axion-degree-freedom}). In co-moving coordinates,
$u^{0}(x)\equiv 1$, $\vec{u}(x) \equiv 0$. Choosing $\Sigma$ to be a hyperplane
at fixed time (in co-moving coordinates), and assuming that the temperature
$T$ is independent of $x=(\vec{x},t)$, (\ref{eq:32}) becomes
\begin{equation}
  \vec{B}({x})\cdot\vec{\nabla}\mu_5(x)=\vec{B}({x})\cdot\vec{\nabla}\dot{\theta}_5(x)\equiv 0,
\label{eq:34}
\end{equation}
which is Eq.~(\ref{eq:12})

\begin{figure}[t]
  \centering
  \includegraphics[width=.5\textwidth]{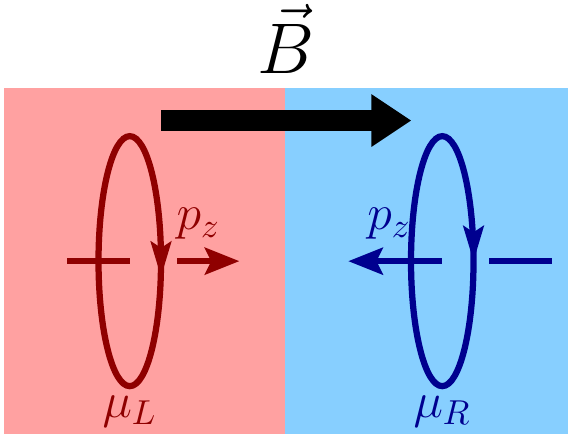}%
  ~\includegraphics[width=.5\textwidth]{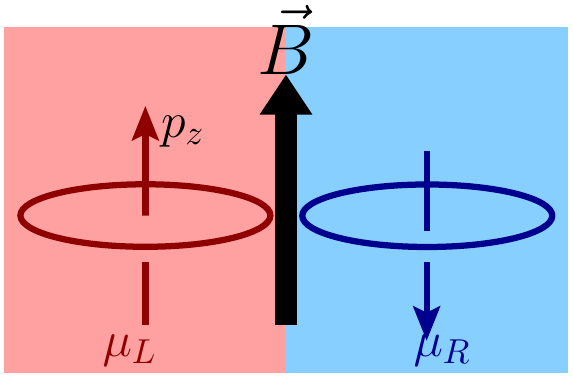}
\caption{The simplest configuration with a non-zero gradient of $\mu_5(x)$. The
  region $x^{1}<0$ has only left-handed particles and the
  region $x^{1}>0$ has only right-handed particles. If we consider a configuration where the magnetic field $\vec B$
has a non-vanishing 1-component, $B_1$, the left-handed particles drift to the right, 
while the right-handed particles drift to the left,
  (because their state is parametrized by a non-zero $p_{1}$). Thus, the state is not on equilibrium state. 
  But if $\vec B$ is perpendicular
  to the $x$-axis the particles move parallel to the
  boundary plane $\lbrace x=0\rbrace$, and the state is an equilibrium state.}
  \label{fig:condition}
\end{figure}

It may be somewhat surprising that the constraint (\ref{eq:34}) must be
imposed on the choice of $\mu_5$. In order to
illustrate the origin of this constraint, we consider a plasma in a homogeneous magnetic field
$\vec{B}\not= 0$. We choose an initial state where all left-handed (spin parallel
to momentum) charged particles are located in the half-space $x^1<0$, while
the right-handed ones (spin anti-parallel to momentum) are located at
$x^1>0$. We assume that the particles are non-interacting. Then the
one-particle states are labelled by the quantum number of a \emph{Landau
  level} and the component of the momentum parallel to $\vec{B}$. If
$\vec{\nabla}\mu_5$ is perpendicular to the field $\vec{B}$ then the densities
of left-handed and right-handed particles are independent of time. But if
$\vec{B}\cdot\vec{\nabla}\mu_5 \neq0$ then left-handed and right-handed
particles start to mix, as they move in the direction of $\vec{B}$. As a
result, the chiral charge density changes in time, which implies that local thermal
equilibrium is lost; see Figure~\ref{fig:condition}.

As a second example, we choose $\mu_5(x)$ to be non-zero and constant in the slab defined by  \mbox{$0\leqslant x^1 \leqslant L$ and $\mu_5(x)\equiv 0$}, elsewhere. 
We consider a $\vec B$-field parallel to the $x^{1}$-axis. An electric current then flows from $x^{1} = 0$ to $x^{1} = L$, thus creating charge densities proportional to $-B\delta (x^{1})$ and $B \delta (x^{1} - L)$, respectively, on the two boundary planes of the slab. These surface charges create a non-vanishing electric field $\vec E (x)$ inside the slab, which is parallel to the $x^{1}$-axis. Hence $\vec E(x) \cdot \vec B(x) \neq 0$, in the region where $\mu_{5}(x) \neq 0$, i.e., inside the slab. By Eq.~(\ref{eq:198}) (Sect. 1.1), $|\mu_{5}|$ then decreases in time, with $\mu_{5}(\vec x, t) \to 0$, as $t \to \infty$. Obviously, local thermal equilibrium is then only approached in the limit where $t \to \infty$.

\subsection{Chiral asymmetry and electro-neutrality}
\label{sec:matter-asymmetry}

To conclude this section, we discuss consequences of electro-neutrality in a high-temperature 
plasma and in the presence of an axion field. Electro-neutrality means that the
\emph{total} electric charge density, $j^{0}_{tot}(x)$, vanishes. 
According to Eq.~(\ref{eq:23}), the light charged particles 
(electrons and positrons, with  $T\gg m_{e}$) make a contribution
\begin{equation}
\label{eq:35}
j^{0}_{\axion} = \frac{\alpha}{\pi}\vec \nabla \theta_{5} \cdot \vec B
\end{equation}
to the electric charge density.  In a plasma containing
\emph{only} electrons and positrons, electro-neutrality implies the
absence of matter-antimatter asymmetry, and $j^0_{\axion}$ must vanish
identically. If $j^0_\axion\equiv 0$ then the 3-divergence of 
$\vec j_\axion$ must vanish, i.e.,
\begin{equation}
  \label{eq:81}
  0 = \vec{\nabla}\cdot \vec
  j_\axion = \frac{\alpha}{\pi}(\vec\nabla \dtheta_5 \cdot \vec B + \vec \nabla \theta_5 \cdot
  \dot{\vec B}) = \frac{\alpha}{\pi}\partial_t(\vec \nabla \theta_5 \cdot
  {\vec B}),
\end{equation}
by \eqref{eq:35}. Expression \eqref{eq:21} for the current density $\vec{j}$ implies 
that $\vec{\nabla}\cdot \vec{j} = 0$, as a consequence of Eq. \eqref{eq:34}. As a
consequence, electro-neutrality, i.e., $j^{0}_{\text{axion}} = 0$ follows.

If the plasma contains several types of charged particles (for example,
$e^\pm$ and $p,\bar p$) then electro-neutrality does $not$ imply any
symmetry between matter and antimatter, and there is no reason to expect that
$j_{\axion}^{0}$, as given in \eqref{eq:35}, vanishes identically! One must then introduce
electric 4-vector current densities for leptons and baryons and axial chemical potentials for all
species of very light, charged particles. For simplicity, we consider a plasma consisting
of electrons, positrons, protons and anti-protons only.
Let $J_{p}^{\mu}$ denote the electric current density of protons and anti-protons, and let
\begin{equation}
\label{eq:360}
j^{\mu}_{p} = \langle J^{\mu}_{p}\rangle_\text{\sc lte}
\end{equation}
Electro-neutrality then implies that
\begin{equation}
\label{eq:370}
j^{0}_{p} = -j^{0}_\axion = -\frac{\alpha}{\pi}\vec{\nabla}\theta_{5} \cdot \vec B
\end{equation}
Let $J^{\mu}_{L}$ be the \emph{total} lepton vector current density, and let $J^{\mu}_{B}$ denote the \emph{total} baryon vector current density. These vector currents are separately conserved; hence  
\begin{equation}
\label{eq:38}
J^{\mu}_{B} - J^{\mu}_{L}
\end{equation}
is \emph{conserved}. In order to describe the matter--anti-matter asymmetry observed in
the universe, we introduce a chemical potential $\mu_\bl =
\dot{\theta}_\bl$ conjugate to $Q_{L} - Q_{B}$, where
\begin{equation}
\label{eq:390}
Q_{L/B} = \int \limits_{t = const}^{} J^{0}_{L/B} (\vec x , t) d^{3}x
\end{equation}
are the conserved lepton and baryon electric charges. If the distribution of asymmetry between matter and 
anti-matter in the universe turned out to be \textit{inhomogeneous} the scalar quantity $\theta_\bl$ would have to be taken to depend on time $and$ space, and a term
\begin{equation}
\label{eq:400}
\int \limits_{\Sigma}^{} u^{\mu}(x) \partial_{\mu}\theta_\bl (x) \left( J^{\nu}_{B}(x) - J^{\nu}_{L}(x)\right)\beta(x)d\sigma_{\nu}(x)
\end{equation}
must be added in the exponent of the expression for the density matrix that
describes local thermal equilibrium, on the right side of Eq.~(\ref{eq:10'});
(see Sect.~\ref{sec:axion-degree-freedom}). To complete the picture we will
have to find an equation of motion for the scalar field $\theta_\bl$. We
note that there are no constraints on $\theta_\bl$ similar to
(\ref{eq:32}), (\ref{eq:34}); but there is an equation analogous to
(\ref{eq:197}) discussed in the next section. Before we turn our attention to
the dynamics of $\theta_\bl$, we study the dynamics of $\theta_{5}$.

\section{Search for the dynamics of the axion field}
\label{sec:dynamics-chiral-asymmetry}
Having introduced a time-dependent axial chemical potential $\mu_5$, it is
necessary to specify its dynamics. Eqs. (\ref{eq:22}) and (\ref{eq:23})
suggest that it may be convenient to derive the dynamics of $\mu_{5}$ from the
one of the axion field $\theta_{5}$.  

In the homogeneous case, i.e., when $\mu_5(t) = \dtheta_5(t)$, Eq.~\eqref{eq:6} can
be rewritten as
\begin{equation}
  \Lambda^2 \overline {\ddot \theta_5} =
  \frac{\alpha}{4\pi} \epsilon^{\mu\nu\lambda\rho}\overline{F_{\mu\nu} F_{\lambda\rho}},
\label{eq:29}
\end{equation}
where $\Lambda$ is a constant with the dimension of an `energy'; (the field $\theta_5$
is dimensionless, because $\mu_5=\dot \theta_{5}$ must be an `energy'). By
Eqs.~(\ref{eq:6}--\ref{eq:198}), we have that
\begin{equation}
  \Lambda^2 =\frac{T^2}6,
\label{eq:300}
\end{equation}
where $T$ is the temperature of the plasma. (We have used that
$\epsilon^{\mu\nu\lambda\rho}F_{\mu\nu} F_{\lambda\rho} = 8 \vec E \cdot \vec
B$, and $\overline{(\cdot )}$ indicates spatial averaging.)  In (\ref{eq:29}),
($F_{\mu\nu}$) is the expectation value of the electromagnetic field tensor in
the state $\langle(\cdot )\rangle_{\text{LTE}}$.  We should ask what the
correct equation of motion of $\theta_5$ is that, upon taking spatial
averages, yields Eq.~(\ref{eq:29}). The axion field
$\theta_5$ has been introduced in order to describe plasmas exhibiting a
chiral asymmetry; in other words, $\theta_5$ may label \emph{states} of a
plasma -- as do other thermodynamic parameters, in particular the
temperature. For this reason, the equation of motion of $\theta_5$ has no
reason to be relativistic; (i.e., to preserve its form under arbitrary
coordinate transformations). It may therefore be plausible to consider a
\emph{diffusion equation} for $\dot{\theta_5}$, which we discuss below.

The chiral anomaly tells us that 
$$
\partial_{\mu}J^{\mu}_{5} = \frac{2 \alpha}{\pi}\vec E \cdot \vec B,
$$
see Eq.~(\ref{eq:6}), Sect~\ref{sec:prev-work}. Taking the expectation in the
state of the plasma (and viewing $\vec E$ and $\vec B$ as classical fields)
yields
\begin{equation}
  \label{eq:41}
  \partial_{\mu}j^{\mu}_{5} = \frac{2 \alpha}{\pi} \vec E \cdot \vec B,
\end{equation}
with $j^\mu_5 = \langle J^\mu_5\rangle_{T,\mu_5}$.
On the basis of Eqs. \eqref{eq:29} and \eqref{eq:41} one argues that
\begin{equation}
  \label{eq:47}
   \partial_{\mu}j^{\mu}_{5} = \Lambda^{2}\ddot{\theta}_{5} + \varepsilon [\theta_{5}], \quad \left(\text{with} \text{   } \Lambda^{2} = \frac{T^{2}}{6}\right),
\end{equation}
where $\varepsilon$ is a term whose spatial average vanishes, i.e.,
$\overline{\varepsilon [\theta_{5}]} = 0$.

If $\theta_{5}$ is interpreted as a kind of thermodynamic parameter labeling
an inhomogeneous state of the plasma in local thermal equilibrium then the following ansatz
for $j^\mu_5$ as a function of $\theta_5$ in co-moving coordinates is reasonable:
\begin{equation}
  \label{eq:48}
  j_{5}^{0} = \Lambda^{2} \dot{\theta}_{5}, \qquad \vec{ j}_{5} = \Lambda^{2} 
  D  \vec {\nabla}(\dot{\theta}_{5}),
\end{equation}
and
\begin{equation}
\label{eq:350}
\Lambda^{2} = \left[\frac{\partial
    j^0_5 }{\partial \mu_5}\right]\biggr|_{\mu_5=0}, \qquad \text{with} \qquad  j^{0}_{5}\vert_{\mu_5 =0} = 0.
\end{equation}
Note that these equations imply that 
\begin{equation}
 \label{eq:36}
  D\vec{\nabla} j^0_5 =  \vec \jmath_5\;,
 \end{equation}
 a relation that can also be confirmed by direct computations;
(see e.g.~\cite{Burnier:12a}). 
In co-moving coordinates, (\ref{eq:41}) and (\ref{eq:48}) then yield the equations of motion
\begin{equation}
  \label{eq:49}
  \Lambda^{2}\left( \ddot{\theta}_{5} - D\Delta \dot{\theta}_{5}\right) = \partial_{\mu}j^{\mu}_{5} = \frac{2 \alpha}{\pi}\vec E \cdot \vec B, 
\end{equation}
which imply (\ref{eq:47}). Together with Eq.~(\ref{eq:24}) we find that
\begin{equation}
  \label{eq:50}
  \Lambda^{2}\left( \dot{\mu}_{5} - D\Delta \mu_{5}\right) = \frac{2 \alpha}{\pi}\vec E \cdot \vec B,
\end{equation}
 in accordance with Eq. (\ref{eq:198}), which is an \emph{inhomogeneous diffusion equation} for $\mu_{5}$. 
This diffusion equation can also be
obtained from a more formal \emph{first order expansion} of the {constitutive
equation} for the global current $j^\mu_5$; (see
e.g.~\cite{Israel:79,Arnold:00}, or the book~\cite{Landau-vol6}).
Taking into account \emph{chirality flips due to small masses} of the charged particles, Eq.~(\ref{eq:50}) should be generalized to the equation 
\begin{equation}
  \label{eq:51}
  \Lambda^{2}\left( \dot{\mu}_{5} - D\Delta \mu_{5}\right) +\Gamma_{f}\mu_{5} = \frac{2 \alpha}{\pi}\vec E \cdot \vec B .
\end{equation}
Furthermore, local thermal equilibrium imposes the constraint
\begin{equation}
  \label{eq:52}
  \vec \nabla \mu_{5} \cdot \vec B = 0,
\end{equation}
see~\eqref{eq:12}. The electric vector current density is given by
 \begin{equation}
  \label{eq:53}
  j^{0}_{tot} = j^{0}_{\text{axion}} + j^{0}_{p} = \frac{\alpha}{\pi}\vec \nabla \theta_{5} \cdot \vec B + j^{0}_{p}
  = 0,
\end{equation}
 \begin{equation}
  \label{eq:54}
  \vec j_{tot} = \vec j_{\text{axion}} + \vec j_{p} = \vec j_{\text{axion}} + \vec j_{Ohm} = \frac{\alpha}{\pi}\left( \mu_{5}\vec B + \vec \nabla \theta_{5} \wedge \vec E\right) +\sigma \vec E,
\end{equation}
with $|\vec j_{p}| \ll |\vec j_{l}|$, because protons are much heavier than
electrons and positrons.

Of course, we also require the \emph{Maxwell equations}, with $j^{\mu} = j^{\mu}_{tot}$ given by (\ref{eq:53}) and (\ref{eq:54}). 

Assuming that the baryonic current $\vec j_{p}$ can be incorporated in $\vec j_{Ohm} = \sigma \vec E$, as we have done in (\ref{eq:54}), we have arrived at a \emph{complete system of field equations} for $\theta_{5}, \vec E $ and $\vec B$!

The conductivity (tensor) $\sigma$ must be calculated from transport equations for the plasma. Apparently, the equation of motion for the scalar field $\theta_{\bl}$ that determines the matter-anti-matter asymmetry is not needed here; although it is of considerable interest interest to derive one, (as will be discussed elsewhere).

Equations (\ref{eq:51})-(\ref{eq:54}) are supposed to be valid in co-moving coordinates. In general coordinate systems equations (\ref{eq:52}) and (\ref{eq:53}) must be replaced by
 \begin{equation}
  \label{eq:52'}
  \mu_{5} = u^{\mu}\partial_{\mu}\theta_{5} \approx \dot{\theta}_{5} + \vec u \cdot \vec \nabla \theta_{5}
\end{equation}
 and
  \begin{equation}
  \label{eq:53'}
  j^{0}_{tot} + \vec u \cdot \vec j_{tot} = 0;
\end{equation}
see Eq.~(\ref{eq:24}). 

The analogue of Eq~(\ref{eq:197}) for the field $\theta_\bl$ is the relation
  \begin{equation}
  \label{eq:550}
  \overline{j^{0}_{B} - j^{0}_{L}} = \varepsilon^{2}\overline{\dot{\theta}_\bl}
\end{equation}
where $\overline{(\cdot)}$ denotes spatial averaging, and $\varepsilon$ is a constant with the dimension of an energy. Since $j^{\mu}_{B} - j^{\mu}_{L}$ is a conserved current, it follows that 
  \begin{equation}
  \label{eq:560}
  \overline{\ddot {\theta}_\bl} = 0
\end{equation}
A relativistic equation compatible with (\ref{eq:560}) is
  \begin{equation}
  \label{eq:570}
  \Box \theta_\bl = 0,
\end{equation}
which describes a free, massless scalar field. As in our discussion of the equation of motion of $\theta_{5}$, one could also envisage an equation of the form
  \begin{equation}
  \label{eq:580}
  \dot{\mu}_\bl - D\Delta \mu_\bl= 0,
\end{equation}
with $\mu_{\bl} = \dot \theta_\bl$ in co-moving coordinates. Choosing the hyperplane
$\Sigma$ appearing in Eq.~(\ref{eq:400}) to correspond to a constant time $t$
(in co-moving coordinates, with $u^{0} = 1, \vec u = 0$) and imposing the
condition that expression (\ref{eq:400}) is \emph{independent} of $t$ (i.e.,
the matter-anti-matter asymmetry does not change in time, anymore, in the old universe), we find the
constraint
  \begin{equation}
    \overline{\left( J^{\alpha}_{B} - J^{\alpha}_{L}\right)\partial_{\alpha} \mu_\bl} = 0.
\end{equation}

\subsection{Dynamics of a fundamental axion field}
\label{sec:action-principle}

It is advisable to also consider the possibility that $\theta_5$ is a 
fundamental field evolving in time according to some relativistic Hamiltonian
dynamics.  The equations of motion for $\theta_{5}$ can then be
derived from an action principle, the action of $\theta_{5}$ being denoted by
\begin{equation}
  \label{eq:420}
  S_{eff}[A, \theta_{5}],
\end{equation}
where $A$ is the electromagnetic 4-vector potential. In order for the action
(\ref{eq:420}) to reproduce the desired expressions for
$j^{\mu}_{\text{axion}}$ given in (\ref{eq:22}--\ref{eq:23})
(Sect.~\ref{sec:axion-degree-freedom}) , it \emph{must} have the form
\begin{equation}
  \label{eq:43}
  S_{eff}[A, \theta_{5}] = S_{0}[\theta_{5}] + \frac{\alpha}{4 \pi}\int d^{4}x\theta_{5}(x) \varepsilon^{\mu \nu \rho \sigma}F_{\mu \nu}F_{\rho \sigma}.
\end{equation}
The equation of motion for $\theta_{5}$ derived from (\ref{eq:43}) is given by
\begin{equation}
  \label{eq:44}
  \frac{\delta S_{eff}[A, \theta_{5}]}{\delta \theta_{5}} = 0,
\end{equation}
or
\begin{equation}
  \label{eq:45}
   \frac{\delta S_{eff}[A, \theta_{5}]}{\delta \theta_{5}} - \frac{\delta S_{0}[\theta_{5}]}{\delta \theta_{5}} = \frac{\alpha}{2 \pi}\varepsilon^{\mu \nu \rho \sigma}F_{\mu \nu}F_{\rho \sigma} = \frac{2 \alpha}{\pi}\vec E \cdot \vec B
\end{equation}
By (\ref{eq:41}), this implies that 
\begin{equation}
  \label{eq:46}
  \frac{\delta S_{eff}[A, \theta_{5}]}{\delta \theta_{5}} - \frac{\delta S_{0}[\theta_{5}]}{\delta \theta_{5}} = \partial_{\mu}j^{\mu}_{5}
\end{equation}
A reasonable ansatz for $S_{eff}[A, \theta_{5}]$ is
\begin{equation}
  S_{eff}[A,\theta_5]=\int d^4x\left\{ \Lambda^{2}(\partial_{\mu}\theta_{5})(x)(\partial^{\mu}\theta_{5})(x) + U(\theta_{5}(x)) - \frac{\alpha}{4 \pi}\theta_{5}(x)\varepsilon^{\mu \nu \lambda \rho}F_{\mu \nu}(x)F_{\lambda \rho}(x)\right\}
\label{eq:31}
\end{equation}
This action yields the equations of motion
\begin{equation}
  \Lambda^2\square\theta_5 + U'(\theta_5)=\frac{\alpha}{4\pi}\epsilon^{\mu\nu\rho\sigma}F_{\mu\nu}F_{\rho\sigma}\;.
\label{eq:30}
\end{equation} 
If $\theta_5$ is spatially homogeneous and $U=0$ this field equation reduces
to~\eqref{eq:29}.  Eq.~\eqref{eq:30} has the form~\eqref{eq:47}, provided
$\overline{U'(\theta_{5})} = 0$; (e.g., if $U \equiv 0$).  Eq.~(\ref{eq:30}),
with a term $\Gamma_{f}\dot{\theta}_{5} $ added on the left side, represents
an alternative to Eq.~(\ref{eq:51}). \\ 
We conclude that, in a system with a fundamental axion field, 
the chiral MHD description of the plasma can be expected to be
applicable \textit{even} if the fermions are not ultra-relativistic. For example,
if an axion-like particle exists and represents a contribution to dark matter, then the production
of dark matter in such a system may be accompanied by the generation of magnetic
fields (see for example~\cite{Schubnel:10}). The phenomenological viability of this scenario will be
discussed elsewhere.

If Eq.~(\ref{eq:30}) is the correct equation of motion it is temping to argue
that not only $F_{\mu\nu}$ corresponds to a quantized field, but the
\emph{axion field} $\theta_5$ must be quantized, too! Furthermore,
the action (\ref{eq:31}), with $U=0$, suggests to unify the electromagnetic
gauge potential A and $\theta_5$ to a gauge potential,
$A=(A,\Lambda\theta_5)$, on a slab of width $\propto \Lambda^{-1}=const\,
\beta$ in \emph{five-dimensional} space-time, with $\beta$ the inverse
temperature; (see (\ref{eq:300})). Then the term $\int d^4x\left\{
  \Lambda^{2}(\partial_{\mu}\theta_{5})(x)(\partial^{\mu}\theta_{5})(x)\right\}$
originates from a contribution to the five-dimensional Maxwell term, after dimensional reduction. 
These ideas are discussed in more detail in Appendix \ref{sec:5d} .

\section{Dynamics of chiral MHD in the inhomogeneous case}
\label{sec:one-mode-solution}

It was demonstrated in~\cite{Boyarsky:11a} that the presence of a dynamical
field $\mu_5(t)$ in the Maxwell-axion equations leads to the generation of 
an ``inverse cascade'' -- i.e., to the transport of
magnetic energy and helicity from shorter to longer scales.  In what follows
we present a preliminary analysis of the inhomogeneous equations derived in
Sect.~\ref{sec:dynamics-chiral-asymmetry} and show that the behaviour of 
solutions is qualitatively similar to that of spatially homogeneous solutions, as
discussed in~\cite{Boyarsky:11a}. In our analysis, we neglect 
\begin{compactenum}
\item \emph{the diffusion term $\Delta \mu_{5}$} in Eq.~\eqref{eq:51},
  assuming that the timescales are much longer than typical diffusion times;
\item the  coupling to the velocity field $\vec{u}$ of the plasma, as in
  Eqs.~\eqref{eq:52'}--\eqref{eq:53'}; i.e., we set $\vec{u} \approx 0$; we also neglect, as usual,
  Maxwell's displacement current proportional to $\dot{\vec{E}}$;
\item the chirality flipping term in Eq.~\eqref{eq:51}, i.e., we set
  $\Gamma_{f} = 0$.
 \end{compactenum}
 We thus analyze the following simplified system of equations:
\begin{equation}
\label{eq:40}
\begin{aligned}
  \vec\nabla \wedge \vec B =& \text{ } \sigma \vec E +
  \frac{\alpha}{\pi}\Bigl(\dot \theta_5 \vec B
  + \nabla \theta_5 \times \vec E\Bigr),\\
  \Lambda^{2} \ddot{\theta_5} =& \text{   }\frac{2\alpha}\pi \vec E\cdot \vec B,\\
  \indent \vec B \cdot \nabla \dot\theta_5 = & \text{ } 0,
\end{aligned}
\end{equation}
with $\mu_5 = \dtheta_5$.

\subsection{Exact single-mode solution}
\label{sec:exact-single-mode}
In this section, we set out to find the simplest (but non-trivial) exact solutions
of Eqs.~\eqref{eq:40}. These solutions can be considered to be direct
generalisations of the ``single-mode'' solutions, considered in
Ref.~\cite{Boyarsky:11a}, to the inhomogeneous case.

We start with the following ansatz for the magnetic induction $\vec{B}$:
\begin{equation}
  \label{eq:42}
  \vec B(x) = B(t)\bigl(\sin(kz),\cos(kz),0\bigr)
\end{equation}
For this choice of $\vec{B}$, we have that
\begin{equation}
  \label{eq:310}
  \vec\nabla\wedge \vec B = k \vec B,
\end{equation}
which is a special case of the \emph{force-free configuration} first
described in~\cite{Lust:54}; (see
also~\cite{Chandrasekhar:56,Chandrasekhar:57}).  A general force-free
configuration~\eqref{eq:310} is given by the following
expression~\cite{Chandrasekhar:57}:
\begin{equation}
\label{eq:122}
\vec B = k^{-1} \vec\nabla\wedge^2(\vec e \psi) + \vec\nabla\wedge(\vec e \psi)
\end{equation}
where $\vec e$ is an arbitrary unit vector and $\psi$ is a solution of the
Helmholtz equation:
\begin{equation}
\label{eq:137}
  \Delta \psi + k^2 \psi = 0.
\end{equation}
The configuration~\eqref{eq:42} corresponds to the choice $\vec e = (1,0,0) $ and $\psi = \sin(kz)$. 
Therefore, a \emph{tracking solution}, found in~\cite{Boyarsky:11a},
\begin{equation}
  \label{eq:121}
  \left\{
  \begin{aligned}
    \dtheta_5 = &\mu_5 =\frac{\alpha}\pi k\\
    \vec E = & 0\\
    \vec B = & B_0\bigl(\sin(kz),\cos(kz),0\bigr)
  \end{aligned}\right.\;,
\end{equation}
is also a solution of the inhomogeneous Maxwell equations~\eqref{eq:40}
\emph{for any value of the constant} $B_0$.  However, if one starts with
non-zero $\vec E(0)$ \emph{or} with $\mu_5 \neq \frac{\alpha}\pi k $ (or if
the initial conditions are not monochromatic, see the next subsection) the
value $B_\infty$, to which the solution $B(t) $ arrives at late times, will
depends non-trivially on the initial conditions as we demonstrate below.

By the Bianchi identity, the electric field corresponding to the
configuration~\eqref{eq:42} is given by
\begin{equation}
  \label{eq:37}
  \vec E = -\frac1{k}\dot B(t) (\sin(k z), \cos(kz),0),
\end{equation}
hence
\begin{equation}
  \label{eq:39}
  \vec E(x)\cdot \vec B(x)  = -\frac1 {2k}\frac{\partial}{\partial t} B^2(t)
  \quad \text{is space-independent}. 
\end{equation}
Next, we plug this ansatz into the system of
Eqs.~\eqref{eq:40}. This yields
\begin{equation}
  \label{eq:61}
  \Lambda^2\pfrac{\mu_5}t = -\frac{\alpha}{2\pi k }\frac{\partial  B^2(t)}{\partial t}
\end{equation}
or
\begin{equation}
  \label{eq:62}
  \mu_5(t) = \mu_5^0 - \frac\alpha{2\pi}\frac{B^2(t) - B_0^2}{k \Lambda^2}
\end{equation}
where $B_0=B(0)$ and $\mu_5^0=\mu_5(0)$.  This is the \emph{exact solution
  for $\mu_{5}(t)$, for all times $t$,} derived from the one-mode ansatz~\eqref{eq:42}. Such a
solution has been first described in~\cite{Boyarsky:11a}, and we see here that
the local nature of the axial anomaly equation~\eqref{eq:40} does not invalidate
the conclusion in~\cite{Boyarsky:11a}. 

Eqs.~\eqref{eq:40} allow us to determine $B(t)$ and its
asymptotic behaviour for large $t$. This is seen by plugging expression~\eqref{eq:62} for $\mu_5(t)$
into the Maxwell equations, which then yield a closed system of equations for $\vec {B}$:
\begin{equation}
  \label{eq:63}
  \vec\nabla \wedge \vec B = \sigma \vec E + \frac\alpha\pi\left( \mu_5^0 -
    \frac\alpha\pi    \frac{B^2(t) - B_0^2}{2k \Lambda^2}\right) \vec {B}
\end{equation}
Using equation~\eqref{eq:37}, and recalling that $\vec\nabla\wedge \vec{B} = k \vec{B}$, 
for $\vec{B}$ as in~\eqref{eq:42}, we arrive at a \emph{non-linear ODE}
  for \textit{one} scalar function $B(t)$:
\begin{equation}
  \label{eq:64}
  k B(t) = -\frac \sigma k \dot B(t) + \frac\alpha\pi\left( \mu_5^0 -\frac\alpha\pi
    \frac{B^2(t) -B_0^2}{2k \Lambda^2}\right) B(t)
\end{equation}
This equation can be integrated explicitly and yields $B(t)$ as a function of
$B_0$ and $\mu_5^0$ and we can write down the explicit time-dependent solution
of the system~\eqref{eq:40}:
\begin{equation}
  \label{eq:206}
\left\{  \begin{aligned}
    B(t) = &\frac{C_1}{\sqrt{1 + C_2 \exp\left({\frac{2 (k^2-\gamma^2)}\sigma
            t}\right)}},\\
    \mu_5(t) =& \mu_5^0 - \frac\alpha{2\pi}\frac{B^2(t) - B_0^2}{k \Lambda^2}
  \end{aligned}\right.
\end{equation}
where we have introduced a parameter $\gamma$ given by
\begin{equation}
  \label{eq:209}
  \gamma^2 := \frac{\alpha}\pi k\mu_5^0 + \frac{\beta_0^2}2, \qquad \text{with}\qquad
  \beta_0^2 := \frac{\alpha^2 B_0^2}{\pi^2 \Lambda^2}. 
\end{equation}
The constants $C_1,C_2$ can be expressed in terms of $B_0,\mu_5^0,
\text{and}\ k$, we will not provide their explicit form. 

Next, we analyse asymptotic behaviour of $B(t)$ as $t\to\infty$. To this end
we put $\dot B(t) = 0$ and find 
an algebraic equation for $B_\infty\equiv B(t\to\infty)$:
\begin{equation}
  \label{eq:208}
  k B_\infty = \frac\alpha\pi\left( \mu_5^0 -\frac\alpha\pi
    \frac{B^2_\infty -B_0^2}{2k \Lambda^2}\right) B_\infty.
\end{equation}
There are two distinct solutions of Eq.~\eqref{eq:208}. The trivial one, i.e.,
$B_\infty = 0$, is approached iff
\begin{equation}
  \label{eq:207}
  \text{Trivial solution} \Leftrightarrow k^2 > \gamma^2\quad\text{or}\quad k > \frac{\alpha \mu_5^0}{2\pi} + \sqrt{\frac{\beta_0^2}2
    +  \parfrac{\alpha \mu_5^0}{2\pi}^2}
\end{equation}
Notice, that although $B_\infty = 0$, the $\mu_5(t\to\infty)$ can still be
non-trivial. In the case of zero magnetic field the asymptotic form of the
$\mu_5(t)$ is not bound to be equal to $\frac{\pi k}\alpha$ and its value is
determined by the Eq.~\eqref{eq:62}.

The non-trivial solution of Eq.~\eqref{eq:65} exists if $k < \gamma$. It is
given by
\begin{equation}
  \label{eq:65}
  B_\infty^2 =
  B_0^2\left[1+\parfrac{2k^2}{\beta_0 ^2
    }\left(\frac{\mu_5^0}{\mu_5^\infty}-1\right)\right],\quad\text{where}\quad
  \mu_5^\infty \equiv \mu_5(t\to\infty) = \frac{\pi k}\alpha,\qquad 
\end{equation}
The necessary condition for the non-trivial solution~\eqref{eq:65} to exist,
i.e. $B_\infty^2 > 0$, is given by
\begin{equation}
  \label{eq:119}
  \left[1+\parfrac{2k^2}{\beta_0 ^2
    }\left(\frac{\mu_5^0}{\mu_5^\infty}-1\right)\right] > 0
\end{equation}
It is exactly opposite to the condition~\eqref{eq:207}, as it should be.

\emph{In summary}, we were able to find an \emph{exact} solution of the
non-linear system of differential equations~\eqref{eq:40} and found
condition~\eqref{eq:207} (or equivalently~\eqref{eq:119}) that relates the
initial value of $\mu_5(t=0)$, $B(t=0)$ and the wave-number $k$ and determines
the ultimate fate of the single-mode solution.

Notice that if, initially, $\mu_5^0 = \frac\pi\alpha k =
\mu_5^\infty$ then the solution is stationary, i.e., $B_0 = B_\infty$.  If,
however, $\frac\alpha\pi \mu_5^0 \neq k$ then $b(t)$ will increase or decrease
and hence cause $\mu_5(t)$ to change in time, as is seen from
Eq.~\eqref{eq:62} .  Indeed, if $\frac\alpha\pi \mu_5^0 >k$ then $b(t) >0$,
and the amplitude of the $\vec{B}$-field grows, while, according
to~\eqref{eq:62}, $\mu_5(t)$ \textit{decreases} -- as it should. But if
$\frac\alpha\pi \mu_5^0 <k$ then $b(t) <0$, i.e., the amplitude of the
$\vec{B}$-field decreases, and, according to~\eqref{eq:62}, $\mu_5(t)$
\textit{increases} towards~$\mu_{5}^{\infty}$.  To demonstrate this, we
consider the behaviour of the solution for small times when $B(t) \approx B_0$
and $\mu_5(t) \approx \mu_5^0$. To this end we linearize the system of
Eq.~\eqref{eq:64} around some special solutions.  We set $B(t) = B_0(1 +
\epsilon b(t))$ where $b(0)=0$. To first order in $\epsilon$ we obtain the
following equation for $b(t)$:
\begin{equation}
  \label{eq:112}
  \left(\frac\alpha\pi \mu_5^0 -k\right) = \left[\left(\frac\alpha\pi \mu_5^0 -k\right) b(t) -\frac{\beta_0 ^2 }{k} b(t)-\frac {\sigma }{k
    } b'(t)\right] + \CO(\epsilon).
\end{equation}
The solution of Eq.~\eqref{eq:112} is given by
\begin{equation}
  \label{eq:210}
  b(t) = \frac{1- \text{exp}\left({\frac{t (\Delta \mu  k-\frac{\beta_0 ^2}2)}{\sigma
        }}\right)}{1-\frac{\beta_0 ^2}{2k \Delta \mu}},\quad\text{where}\quad \Delta\mu = \frac{\alpha \mu_5^0}{\pi}-k.
\end{equation}
and it demonstrates the conclusion,

\subsection{Stability analysis of the helical single-mode solution}
\label{sec:gener-line-pert}

In this section we sketch an analysis of stability of the solution found in
the previous section. To this end, we assume that the initial configuration
has reached its time-independent form with some $k$, $B_\infty$ and
$\mu_5^\infty$. We then perturb the vector potential $\vec A$ and the chiral
chemical potential $\mu_5$ by
\begin{equation}
  \label{eq:115}
  \begin{aligned}\
    \mu_5(\vec{x},t) =& \frac{\pi k}{\alpha} + \delta \mu_5(\vec{x},t)\\
    \vec A(\vec{x},t) = & \frac {\vec B_0(z)}k +  \vec a (\vec{x},t)
  \end{aligned}
\end{equation}
where $\vec B_0(z)$ has the form~\eqref{eq:42}, with amplitude $B(t) \equiv B_\infty$
 given by Eq.~\eqref{eq:65}. In \eqref{eq:115}, $\mu_{5}(\vec{x},t) \equiv \mu_{5}^{\infty}$ if
 $\delta \mu_5(\vec{x},t) \equiv 0$; see \eqref{eq:65}.

 The (equation expressing the) chiral anomaly (see Eq.~\eqref{eq:40}),
 linearized around the background $\mu_{5}(x)\equiv \alpha^{-1}{\pi} k$ and
 $\vec{A}(x) \equiv k^{-1}\vec{B}_{0}(z)$, then yields

\begin{equation}
  \label{eq:500}
 \Lambda^{2} \pfrac{\delta \mu_5}{t} = -\frac{2\alpha}\pi \vec B_0(z)\cdot \pfrac{
    \vec a}{t},
\end{equation}
which can be integrated to
\begin{equation}
  \label{eq:183}
  \delta \mu_5(\vec{x},t) = -\frac{2\alpha}{\pi} \frac1{\Lambda^2} \vec B_0(z)\cdot 
  \Bigl( 
  {\vec a(\vec{x},t)} -  {\vec a_0(\vec{x})}\Bigr) + \delta \mu_5(\vec{x},0)
\end{equation}
where $\vec a_0(\vec{x}) := \vec a(\vec{x},t{=}0)$.  Below, we choose as \textit{initial condition}
 $\delta \mu_5(\vec{x},t=0) = 0$. This is obviously a special choice.
However, physically, it is the most interesting one, because the spatial inhomogeneity of
the chiral chemical potential is then caused by fluctuations of the magnetic
field.  The analysis of a more general situation will follow. Using~\eqref{eq:183},
the equation of motion for $\vec a$ is seen to be given by
\begin{equation}
  \label{eq:188}
  -\Delta \vec a = -\sigma\pfrac{ \vec a}t + k\vec\nabla\wedge (\vec a)
  -\frac{2\alpha^2}{\pi^2} \frac1{\Lambda^2} \Bigl( \vec B_0(z)\cdot {\vec a} - \vec B_0(z)\cdot \vec a_0\Bigr) \vec B_0(z).
\end{equation}
We analyse the special solution found by assuming that $\vec a$ depends only
on the spatial coordinate $z$ and on time $t$; i.e., $ \vec a(\vec{x},t) =
\bigl(a_x(z,t), a_y(z,t), 0\bigr)$. The equations then reduce to
\begin{equation}
  \label{eq:146}
  \begin{aligned}
    a_x''& = \sigma \dot a_x + k a'_y + 2\beta^2 \bigl(\sin^2(kz) (a_x-a_x^0)
    + \sin(kz)
    \cos(kz) (a_y-a_y^0) \bigr),\\
    a_y''& = \sigma \dot a_y - k a'_x + 2\beta^2 \bigl (\sin(kz) \cos(kz)
    (a_x-a_x^0) + \cos^2(kz) (a_y-a_y^0) \bigr),
  \end{aligned}
\end{equation}
where $\beta^{2}$ is given by
\begin{equation}
  \beta^2 \equiv \frac{\alpha^2 B_\infty^2}{\pi^2 \Lambda^2}.  \label{eq:120}
\end{equation}
(similar to the definition of $\beta_0$ in~\eqref{eq:209}, but with $B_0$
replaced with $B_\infty$).

We rewrite the system of equations \eqref{eq:146} as a matrix equation
\begin{equation}
  \label{eq:157}
  \underbrace{\begin{pmatrix}
    \sigma \partial_t + \beta^2 - \partial_z^2 & k \partial_z\\
    -k \partial_z & \sigma \partial_t + \beta^2 - \partial_z^2
  \end{pmatrix}}_{\equiv \hat L}
  \begin{pmatrix}
    a_x\\
    a_y
  \end{pmatrix}= -\beta^2\underbrace{\begin{pmatrix}
      -\cos(2 kz) & \sin(2 k z)\\
      \sin(2 kz) & \cos(2 kz)
  \end{pmatrix}}_{\equiv O(z)}
\begin{pmatrix}
    a_x-a_x^0\\
    a_y-a_y^0
  \end{pmatrix},
\end{equation}
or, schematically, as $\hat L \vec a = -\beta^2 O(z) (\vec a-\vec a_0)$.
Defining $U$ by
\begin{equation}
  \label{eq:160}
  U =
  \begin{pmatrix}
    1 & 1\\
    -i & i
  \end{pmatrix},
\end{equation}
we find that
\begin{equation}
  \label{eq:164}
  U^{-1} \hat L U = \begin{pmatrix}\sigma \partial_t + \beta^2 - \partial_z^2
    -i k \partial_z & 0
    \\
    0 & \sigma \partial_t + \beta^2 - \partial_z^2
    +i k \partial_z
\end{pmatrix}=\begin{pmatrix}
  \hat L_+&0\\
  0 & \hat L_-
\end{pmatrix},
\end{equation}
and
\begin{equation}
  \label{eq:166}
  U^{-1} \begin{pmatrix}
    a_x\\
    a_y
  \end{pmatrix} = \frac12 \begin{pmatrix}
    a_x+i a_y\\
    a_x -ia_y
  \end{pmatrix}=
\begin{pmatrix}
  a_+\\
  a_-
  \end{pmatrix}.
\end{equation}
Furthermore,
\begin{equation}
  \label{eq:167}
  U^{-1} O(z) U = \begin{pmatrix}
    0 & - e^{-2i kz}\\
    -e^{2i kz} & 0
  \end{pmatrix}.
\end{equation}
We then have that
\begin{equation}
  \label{eq:189}
    \begin{aligned}
      \hat L_+ a_+ &= \beta^2 e^{-2i kz} \bigl(a_- - a_-^0\bigr),\\
      \hat L_- a_- &= \beta^2 e^{+2i kz} \bigl(a_+ - a_+^0\bigr),
  \end{aligned}
\end{equation}
and we choose as an initial condition a plane wave: $a_-^0 = e^{iqz}$,
$a_+^0=e^{-i q z}$, hence $\vec\nabla\wedge\vec a_0 = q \vec a_0$.  One may attempt to
solve Eqs.~\eqref{eq:189} by iteration. In zeroth approximation, we set the right
side of Eqs.~\eqref{eq:189} to zero and get
\begin{equation}
  \label{eq:190}
  \begin{aligned}
    \hat L_+ a_+^\0 &= 0 ; & a_+^\0(0,z) &= e^{-iqz},\\
    \hat L_- a_-^\0 &= 0 ; & a_-^\0(0,z) &= e^{iqz}.\\
  \end{aligned}
\end{equation}
These equations are solved by
\begin{equation}
  \label{eq:191}
  \begin{aligned}
    a_+^\0 = e^{-iqz}e^{\lambda t},\quad a_-^\0 = e^{iqz}e^{\lambda
      t},\qquad \lambda = \frac{kq - q^2 -\beta^2}{\sigma}.
  \end{aligned}
\end{equation}
The condition that
perturbation grows in time, i.e. $\lambda > 0$, is translated into
\begin{equation}
  \lambda > 0 \Longleftrightarrow \frac{k - \sqrt{k^2 -4\beta^2}}2< q< \frac{k + \sqrt{k^2 -4\beta^2}}2
\label{eq:194}
\end{equation}
That is, for $\beta^2 < k^2/4$ and $q$ obeying the inequalities
in~\eqref{eq:194}, a perturbation with wavenumber $q$ grows in time; (i.e.,\
the single-mode solution of Section~\ref{sec:one-mode-solution} is unstable).
We stress that, compared to the results of Ref.~\cite{Boyarsky:11a}, not every
mode, longer than $2\pi/k$, can grow. The difference is given by the presence
of $\beta^2$ term, which is a non-linear contribution from the background
field, $B_\infty$. For $\beta^2 > k^2/4$, the solution of the previous
Section~\ref{sec:exact-single-mode} is \emph{stable} with respect to the
perturbation in the form

To point out an important difference between the stability analysis of
Eqs.~\eqref{eq:40} and the results of~\cite{Boyarsky:11a}, we iterate
Eqs.~\eqref{eq:189} again, yielding
\begin{equation}
  \label{eq:192}
  \begin{aligned}
    \hat L_+ a_+^\1 &= \beta^2 e^{-i(2k-q)z}(e^{\lambda t} -1)  ; & a_+^\1(0,z) &= 0,\\
    \hat L_- a_-^\1 &= \beta^2 e^{i(2k-q)z}(e^{\lambda t} -1) ; & a_-^\1(0,z) &= 0,\\
  \end{aligned}
\end{equation}
which is solved by  (we only write the explicit expression for $a_+^\1$, $a_-$ being
its complex conjugate)
\begin{small}
  \begin{equation}
    \label{eq:193}
    a_+^\1 =   \frac{\beta^2 e^{-i(2k-q)z}}{2k(k-q)} \left[ (e^{\lambda
        t}-1) + \frac\lambda{\lambda_1}(e^{-\lambda_1 t} -1)\right], \qquad \lambda_1 =
    \frac{2k^2 - 3k q + q^2 + \beta^2}{\sigma} > 0,
  \end{equation}
\end{small}
for $q<k$.  We see that $a_+^\1$ is a wave with inverse wavelength $2k-q$. If
$q<k$ (i.e., at $t=0$, the perturbation has a longer wavelength than the
original single-mode solution) then $2k -q > k$; i.e.,\ a wave with shorter
wavelength is excited. This result is a consequence of the non-linear nature
of Eqs.~\eqref{eq:40} and the coupling between the modes (i.e. the generation
of mode $2k -q$ starting from two modes with the wave-number $k$ and $q$) is
expected.

Below we show that the coefficient in front of the $e^{\lambda t}$ term in
$a^\1_+$ is always smaller than the term in front of the $e^{-i qz}$
harmonic. The ratio of coefficients can be found to be (we take times large
enough so that $\lambda_1 t \gg 1$ and $\lambda t\gg 1$)
\begin{equation}
  \label{eq:195}
  \frac{|a^\1_+(t,z)|}{|a^\0_+(t,z)|}\biggr|_{\lambda_1 t \gg 1;\lambda t \gg 1}=
  \frac{\beta^2}{2k(k-q)} \le \frac{q}{2k} < \frac1{2}
\end{equation}
(in view of the condition that $\beta^2 < k q - q^2$ that follows from $\lambda
> 0$, Eq.~\eqref{eq:194} and considering only modes with $q < k$). As a result
at most $\frac14$ of the energy could be transformed into the short wavelength
mode.

\paragraph{Summary:} Starting from the simplified
system of equations of chiral electrodynamics, see Eqs.~\eqref{eq:40}, we have shown 
that the helical single-mode solution is stable
for sufficiently strong magnetic fields. For weak
magnetic fields, the long-wavelength perturbation grows in time. A short-wavelength
mode also gets excited, but its amplitude is always parametrically smaller
than the one of the long-wavelength mode. Thus, transfer of energy and
helicity from a single helical field mode to a long-wavelength perturbation, as
described in~\cite{Boyarsky:11a}, is  observed for solutions of the full system of
equations with an inhomogeneous axial chemical potential.

\section{Conclusion and outlook}
\label{sec:conclusion-outlook}

In a relativistic plasma of charged particles, the axial anomaly is accompanied 
by the appearance of new degrees of freedom 
described by a pseudo-scalar field $\theta_5(x,t)$.
This field has the property that its time derivative is equal to the axial chemical potential
$\mu_5$ parameterizing states of local equilibrium of the plasma.  
When coupled to the electromagnetic field and to the motion of the
fluid the field $\theta_{5}$ appears in a system of equations that we call the ``equations
of \emph{chiral magnetohydrodynamics}''; (see Eqs.~\eqref{eq:48} through
\eqref{eq:53}). If $\theta_5$ is spatially uniform and depends only on time
$\dtheta_5$ is nothing but the axial chemical potential $\mu_{5}$ of the
light charged fermions.  In this paper, we have studied states of the plasma corresponding
to \emph{local}
equilibrium, with $\theta_5$ depending not only on time $t$, but
also on $\vec x$. Our analysis is significant, because the presence of a
non-zero (homogeneous) axial chemical potential $\mu_5$ leads to an
instability of some solutions of the Maxwell equations accompanied by
the generation of helical magnetic fields~\cite{Joyce:97,Frohlich:2000en,Frohlich:2002fg}. 
This instability leads to the appearance of strong electromagnetic fields that will, in turn,
back-react on $\mu_5$, making it spatially inhomogeneous.

We have analyzed special solutions of the system of chiral MHD equations and
have shown that the qualitative conclusions reached in~\cite{Boyarsky:11a}
remain valid. In particular, we ave shown that there is a stationary solution
(helical single-mode solution) of our system of non-linear equations with the
property that the growth caused by a non-vanishing axial chemical potential
exactly compensates the Ohmic dissipation
(Section~\ref{sec:one-mode-solution}).  Another important property of
solutions in the presence of a homogeneous axial chemical potential is the
``inverse cascade'' phenomenon, i.e., the transfer of energy and magnetic
helicity from short to large scales, described in
Subsection~\ref{sec:exact-single-mode}. This transfer is not caused by
turbulence in the plasma, but is an effect derived from the chiral anomaly and
caused by the chiral imbalance in the plasma .  We have shown that, under
suitable conditions (Section~\ref{sec:gener-line-pert}), this phenomenon
appears to persist for spatially inhomogeneous fields.

Our analysis can be expected to be relevant for the study of various physical systems 
exhibiting a chiral (left-right) imbalance, including the quark-gluon plasma and 
the plasma in the early Universe. In an analysis of the plasma in the early Universe, 
it might not be legitimate to neglect the coupling of the axion field $\theta_5$ to the space-time
curvature, (namely to the term $\frac12 \epsilon^{\mu\nu\alpha\beta} R_{\mu\nu\,
  \sigma\tau} R_{\alpha\beta}^{~~~\sigma\tau}$, where $R_{\mu\nu\, \sigma\tau}$
is the Riemann tensor). We plan to return to this topic elsewhere.

Many elements of our analysis remain unchanged when the
axion is treated as a fundamental field with relativistic hamiltonian dynamics. 
As discussed in Sect. 4, it can induce a left-right asymmetry in the distribution of 
electrically charged fermions and trigger the growth of magnetic fields.

\subsection*{Acknowledgements}
J.\ F.\ thanks A. Alekseev, V. Cheianov and B. Pedrini for countless
discussions and collaboration on related matters. A.~B. and O.~R.\ thank
A.~Abanov, V. Cheianov and M.~Shaposhnikov for discussions at various stages
of this project. J.~F.\ thanks  IHES for hospitality during various periods
of work on this project.  A.B.\ acknowledges support from the ``Fundamentals
of Science'' program at Leiden University during the last stage of our
efforts.

\bibliographystyle{JHEP-2} %
\bibliography{\jobname}

\appendix

\section{Electrodynamics on a slab in five-dimensional Minkowski space}
\label{sec:5d}

\begin{figure}[t]
  \centering
  \includegraphics[height=4cm]{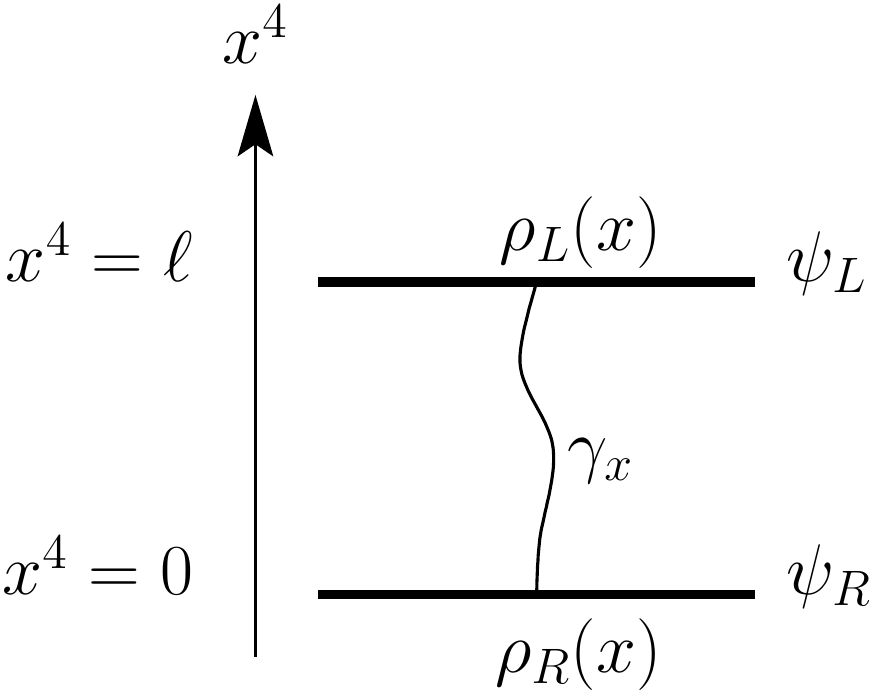}
  \caption{Five-dimensional geometry. Light chiral fermions are localized on the boundaries of the five-dimensional slab ($x^4=0$ and $x^4 = \ell$).}
  \label{fig:5dslab}
\end{figure}

The connection between the chiral chemical potential $\mu_{5}$ and the axion field $\theta_{5}$
becomes manifest if one studies electrodynamics on a slab $\mathbb{R}^{3,1}\times \mathcal{I}$
of \textit{five-dimensional} Minkowski space, where $\mathcal{I}$ is a finite interval extending into
the fifth dimension;
(cf.~\cite{Frohlich:2000en,Frohlich:2002fg,Froehlich:12a}).  Here we briefly review some 
important elements of this story. We consider
very heavy charged four-component Dirac fermions populating the 5-dimensional
``bulk''. After coupling these fermions to an external electromagnetic
5-vector potential $A$ and, subsequently, integrating them out, we obtain a low-energy
bulk effective action for the gauge field $A$ and a boundary effective action for boundary
degrees of freedom, which turn out to be \emph{massless chiral fermions} localized
on the $(3+1)$-dimensional ``top'' and ``bottom'' boundary components 
of the slab (see Fig.~\ref{fig:5dslab}).\footnote{These chiral fermions may acquire a mass
  through tunneling between the two boundary components, which shows that they may
  be used for purposes of reasonably realistic model building. We will ignore this
  possibility in the following.}   The effective action is given by
\begin{eqnarray}
S^{(5)} & = & \int d^5x \, \left[-\frac{1}{4\ell} F_{ab}^2 +
    \frac{\alpha}{32\pi}\epsilon^{abcde}A_a F_{bc}  F_{de}\right]
    \nonumber \\
   & & + \text{ boundary action for chiral fermions}
  \nonumber\\
 & &  + \text{ higher-order non-local terms},
     \label{eq:123}
\end{eqnarray}
where $\ell$ is the length of the interval $\mathcal{I}$, $\alpha$ is the 4-dimensional
fine-structure constant, and $F_{ab}$ is the 5-dimensional field strength
($a,b=0,1,2,3,4$).  Fermions of opposite chirality are located on
opposite boundary components, i.e., at $x^4=0$ and $x^4 = \ell$, respectively. We turn on 
an electric field pointing into the fifth dimension, i.e., $F_{0a}=0$, for $a=1,2,3$, and $F_{04} = const.$.
Then there are non-zero charge densities,
$\rho_L(x)$ and $\rho_R(x)$, of chiral fermions located on opposite boundary components of the slab.
These left- and right-chiral fermions couple to the electromagnetic gauge field restricted
to the boundary hyperplanes. They give rise to anomalous currents 
on both boundary hyperplanes. The
action $S^{(5)}$ in ~\eqref{eq:123} \textit{must} therefore contain the
 \emph{Chern-Simons term} that cancels
the gauge anomalies of the chiral fermions on the two boundary components.
The bulk electric current corresponding to the action $S^{(5)}$ is given by 
$$ j^{a} := \frac {\delta S^{(5)}}{\delta A_{a}}.$$ 
It has a contribution corresponding to the Chern-Simons
term 
\begin{equation}
  \label{eq:124}
  j_\textsc{cs}^a = \frac{\alpha}{8\pi} \epsilon^{abcde} F_{bc} F_{de}
\end{equation}
Note that $j^{a}_{CS}$ is \textit{divergence-free} in the five-dimensional bulk. 
Its divergence does, however, \textit{not} vanish on the boundaries, where it is
proportional to $\pm \frac{\alpha}{8\pi}\epsilon^{\mu\nu\lambda\rho}
F_{\mu\nu}F_{\lambda\rho}$, respectively, and cancels the divergence of the 
anomalous currents of the massless chiral fermions located on the boundary components 
of the slab -- a phenomenon known as \emph{anomaly inflow}.

In order to reveal the connection between this model of 5D electrodynamics 
and axion electrodynamics, we study the dimensional reduction of the five-dimensional theory. 
We define a scalar field
depending only on the coordinates $x=(x^{0}=t, x^{1}, x^{2}, x^{3})$ 
of a space-time point $X\in \mathbb{R}^{3,1}\times \mathcal{I}$ by setting
\begin{equation}
  \label{eq:125}
  \theta_5(x) = \int_{\gamma} dx^4 \, A_4(X)
\end{equation}
where $\gamma$ is a straight path in the fifth direction connecting the two points $X_{l}=(x, x^{4}=0)$ and
$X_{u}=(x, x^{4}=\ell)$ located on opposite boundary components; see Fig.~\ref{fig:5dslab}.
We assume that the components $A_{\mu}, \mu=0,1,2,3$ of the
gauge field $A$ are independent of the $x^4$-coordinate (or are averaged over $x^{4}$). 
By Eq.\eqref{eq:125}
\begin{equation}
  \label{eq:127}
  \dot \theta_5 = -\int_\gamma dx^4\, \dot A_4= \int_{\gamma} dx^{4} E_{4}.
\end{equation}
The right side of this equation is the \emph{voltage drop} between the two boundary components, 
which is nothing but the difference of the chemical potentials of left- and right-chiral fermions.  
The dimensional reduction of the action~(\ref{eq:123}) yields exactly the effective action given in Eq.~\eqref{eq:31} $+$ a 4D Maxwell term. 
This implies that the axion-photon coupling constant can be identified with the size of the
interval $\mathcal{I}$ extending into the fifth dimension. Comparing the equations derived
 in this appendix with Eqs. (5) and (6) of Sect. 1, we observe that the temperature $T$ can be identified
 with the inverse of the width $\ell$ of the slab.
 
 The theory outlined here can be viewed as a five-dimensional cousin of the quantum Hall effect.

\section{Linear analysis}
\label{sec:linear-analysis}

\subsection{Growth of the long-wavelength modes}
\label{sec:growth-long-wavelength}

Let us now perturb the stationary solution for the one-mode~\eqref{eq:63} by a
small long-wavelength mode $\vec B_1$:
\begin{equation}
  \label{eq:540}
  \vec B_1(z,t) = \epsilon B_0 b_1(t) \bigl(\sin(k_1 z),\cos(k_1 z), 0\bigr)
\end{equation}
The linearised equation does not have the simple form Eq.~\eqref{eq:112}
anymore.  However the property~\eqref{eq:39} still holds \emph{for the mode}
$\vec B_1$ separately. What is not true, however is that the overall
$\mu_5(t)$ is now space-independent. It is governed by the following equation
(where again we introduced $B(t) = B_0 \bigl(1+ \epsilon b(t)\bigr)$
\begin{equation}
  \label{eq:55}
  f_5^2 \dot \mu_5 =\frac{2\alpha}\pi \left[\pfrac{}t \frac{ B^2(t)}{2k}+\epsilon^2\pfrac{}t
    \frac{b_1^2(t)}{2k_1} +\epsilon\left(\frac{ B(t) \dot b_1(t)}{2k_1} + \frac{b_1(t) \dot B(t)}{2k}\right)\cos\bigl((k-k_1)z\bigr)\right]
\end{equation}
To the first order in $\epsilon$ Eq.~\eqref{eq:55} has the following form
\begin{equation}
  \label{eq:56}
  f_5^2 \pfrac{ \mu_5}t =-\epsilon\frac{2\alpha}\pi B_0 \pfrac{}t\left[\frac{
      b(t)}{k}+ \frac{
      b_1(t)}{k_1}\cos\bigl((k-k_1)z\bigr)\right]   
\end{equation}
We can integrate Eq.~\eqref{eq:56} over time to get:
\begin{equation}
  \label{eq:58}
  \boxed{\mu_5(t) = \mu_5^0 - \epsilon\frac{2\alpha}\pi \frac{B_0}{f^2_5} \left[\frac{
        b(t)}{k}+ \frac{
        b_1(t)}{k_1}\cos\bigl((k-k_1)z\bigr)\right] }  
\end{equation}
It is important to notice that in this order in $\epsilon$
\begin{equation}
  \label{eq:57}
  \nabla \theta_5 \propto (0,0,1)
\end{equation}
--- points in the $z$-direction and therefore $\vec B\cdot \nabla \theta_5$ is
indeed equal to zero.

If $\mu_5$ is homogeneous in space (ie. if we neglect the last term in
Eq.~\eqref{eq:58}) we end us with the usual two-mode equation:
\begin{equation}
  \label{eq:530}
  \begin{aligned}
    k B(t) & = -\frac{\sigma}{k}\dot B(t) + \frac\alpha\pi\mu_5(t) B(t)\\
    k_1 b_1(t) & = -\frac{\sigma}{k_1}\dot b_1(t) + \frac\alpha\pi\mu_5(t)
    b_1(t)
  \end{aligned}
\end{equation}
If instead of solving equations~\eqref{eq:530} directly we would expand them in
$\epsilon$ (in view of the subsequent non-homogeneous case) we would get the
following. 

We start with the \fbox{$\mu_5^0 = \frac{\pi k}\alpha$} (i.e. the one that
would make a single mode stationary). We then see that the correction to this
mode obeys an ODE (compare the last two terms in Eq.~\eqref{eq:112})
\begin{equation}
  \label{eq:59}
  \sigma b'(t) = -\frac{\alpha^2}{\pi^2}\frac{B_0^2}{f_5^2} b(t)
\end{equation}
This equation has the solution $b(t)=0$ if one starts from $b(0)=0$. This
probably means that $B(t)$ does not change in the linear order in $\epsilon$.
In the first order in  $\epsilon$ the eq. for $b_1(t)$ has the form, expected
from~\eqref{eq:530}:
\begin{equation}
  \label{eq:110}
  \sigma b_1'(t) = (k k_1 - k_1^2)b_1(t)
\end{equation}
which gives exactly the exponential solution with $\mu_5=\frac{\pi k}\alpha$.

Qualitatively one has constant amplitude for the short wavelength solution
(the non-trivial evolution appears at the order $\epsilon^2$). The growth of
the mode $b_1(t)$ starts immediately and goes on until the neglected terms
become important.

\subsection{Two-mode inhomogeneous solution}
\label{sec:two-mode-inhom}

Let us now turn to the case of two-mode inhomogeneous solution. The expression
for $\mu_5$ is given by Eq.~\eqref{eq:58} and the following equations for the
different modes appear:

The mode with the wave vector $k$ (\emph{short wavelength})
\begin{equation}
  \label{eq:111}
  \sigma b'(t) =
  -\frac{\alpha^2}{\pi^2}\frac{B_0^2}{f_5^2} b(t) \quad\text{(identical to\protect~\eqref{eq:59})}
\end{equation}
The mode with the wave vector $k_1$ (\emph{long wavelength}):
\begin{equation}
  \label{eq:113}
  \sigma b_1'(t) = (k k_1 - k_1^2)b_1(t) + \frac{\alpha^2 B_0^2}{2\pi^2f_5^2
  }\Bigl(b_1(t) - b_1(0)\Bigr)
\end{equation}
Due to the non-linearity of the equations, we also have the \emph{very short
  wavelength mode} ($2k-k_1$) excited. If we assume that its initial amplitude
was zero, we have
\begin{equation}
  \label{eq:114}
  {\sigma b_2'(t)} = \frac{\alpha^2 B_0^2({2k-k_1})}{2\pi^2 k_1 f_5^2
  }\Bigl(b_1(t) - b_1(0)\Bigr)
\end{equation}
--- a term, similar to the last term in Eq.~\eqref{eq:113} but with an
additional ``enhancement'' $k/k_1\gg 1$. However, this term is sourced by
$(b_1(t) - b_1(0))$ and will not get excited until the mode $b_1$ had grown
significantly.

\end{document}